\documentclass[12pt]{article}

\usepackage{graphics}
\usepackage{epsfig}
\input epsf

\title{Study of non-collinear parton dynamics in the prompt photon photoproduction at HERA}

\author{A.V.~Lipatov, N.P.~Zotov}

\begin{document}

\maketitle

\begin{center}

{\it Skobeltsyn Institute of Nuclear Physics,\\ 
Lomonosov Moscow State University,
\\119991 Moscow, Russia\/}\\[3mm]

\end{center}

\vspace{0.5cm}

\begin{center}

{\bf Abstract }

\end{center}

We investigate the 
prompt photon photoproduction at HERA 
within the framework of $k_T$-factorization QCD approach.
Our consideration is based on the off-shell matrix elements for the underlying partonic subprocesses.
The unintegrated parton densities in a proton and in a photon are determined using 
the Kimber-Martin-Ryskin (KMR) prescription. Additionally, we use the CCFM-evolved unintegrated 
gluon as well as valence and sea quark distributions in a proton.
A conservative error analisys is performed.
Both inclusive and associated with the hadronic jet production rates
are investigated. The theoretical results are compared with the recent
experimental data taken by the H1 and ZEUS collaborations. 
We study also the specific kinematical properties of the photon-jet system
which are strongly sensitive to the transverse momentum of incoming partons.
Using the KMR scheme, the contribution from the quarks emerging from the earlier steps of the parton 
evolution is estimated and found to be of 15 -- 20\% approximately.

\vspace{0.8cm}

\noindent
PACS number(s): 12.38.-t, 13.85.-t

\vspace{0.5cm}

\section{Introduction} \indent 

The prompt photon production in $ep$ collisions at HERA is  subject of 
intense studies~[1--6]. The theoretical and experimental investigations of such processes
have provided a direct probe of the hard subprocess dynamics, since produced photons 
are largely insensitive to the effects of final-state hadronization.
Usually photons are called "prompt" if they are coupled to the interacting quarks.
From the theoretical point, these photons in $ep$ collisions can be produced
via direct $\gamma q \to \gamma q$ and resolved production mechanisms. In resolved
events, the photon emitted by the electron fluctuate into a hadronic state
and a gluon and/or a quark of this hadronic fluctuation takes part in the hard
interactions. Prompt photon measurements can be used also to constrain the
parton densities in the proton and in the photon.

Recently the H1 and ZEUS collaborations have reported data~[2--6] on 
inclusive and associated (with the hadronic jet) prompt photon production at HERA.
However,  next-to-leading order (NLO) collinear pQCD 
calculations~[7, 8] are $30 - 40$\% below 
these data, especially in rear pseudo-rapidity (electron direction) region. 
It was demonstrated~[2--5] that the observed disagreement is difficult to explain with conventional
theoretical uncertainties connected with scale dependence and 
parametrizations of the parton densities.
The origin of the disagreement has been ascribed to the effect of initial-state 
soft-gluon radiation. It was shown~[3] that observed
 discrepancy can be reduced by introducing some additional intrinsic transverse 
momentum $k_T$ of the incoming partons. The ZEUS fit to the data gave a $k_T$ value of about 1.7~GeV~[3]. 
A similar situation is observed also at Tevatron energies:
in order to describe the measured transverse momentum distributions of the photon 
the Gaussian-like $k_T$ spectrum with an average value of $k_T \sim 3$~GeV was introduced~[9, 10].
Of course, such large partonic $k_T$ must have a significant perturbative QCD
component.

The transverse momentum of incoming partons 
naturally occurs in the framework of $k_T$-factorization approach of QCD~[11].
In this approach, the transverse momentum $k_T$ is generated perturbatively 
in the course of non-collinear parton evolution via the 
corresponding (usually Balitsky-Fadin-Kuraev-Lipatov (BFKL)~[12] or 
Ciafaloni-Catani-Fiorani-Marchesini (CCFM)~[13]) evolution equations.
A detailed description of the $k_T$-factorization can be
found, for example, in reviews~[14--16]. 
As it was demonstrated in the ZEUS paper~[4] and in the recent experimental study~[6] performed
by the H1 collaboration, the $k_T$-factorization predictions~[17] 
for prompt photon photoproduction at HERA are in better
agreement with the data than the published results of the collinear NLO pQCD calculations~[7, 8].

An important component of the first calculations~[17] in
the framework of $k_T$-factorization approach was the 
unintegrated quark distributions $f_q(x,{\mathbf k}_T^2,\mu^2)$ in a proton. 
These quantities are poorly known  
since there are theoretical difficulties 
in obtaining the quark distributions directly from CCFM
equation (see also~[14--16] and references therein for more information). 
At present, the unintegrated quark densities are most often used
in the framework of KMR~[18] approximation only. 
As a result, the dependence of the $k_T$-factorization predictions~[17] 
on the non-collinear evolution scheme has not been investigated.
This dependence in general can be significant and it is a special 
subject of study in the $k_T$-factorization approach. 

Therefore, in the present paper in addition to the KMR approach 
we propose a some simplified way to evaluate the unintegrated 
quark densities $f_q(x,{\mathbf k}_T^2,\mu^2)$ within the CCFM dynamics. 
First we convolute the CCFM-evolved gluon distribution $f_g(x,{\mathbf k}_T^2,\mu^2)$ with the 
usual unregulated leading-order DGLAP splitting function $P_{qg}(z)$
to obtain the unintegrated sea quark densities.
Then we add the CCFM-evolved valence quark densities
which have been recently evaluated and applied~[19] to the jet production at the LHC conditions (in the
framework of Monte-Carlo event generator \textsc{cascade}~[20]).
Of course, in this way we only simulate the last gluon splitting in the full evolution cascade
and do not take into account contribution from quarks
coming from the earlier steps of the evolution. But it is not evident a priori, whether
the last gluon splitting dominates or not. One of the goals of our
study is to clarify this point. In order to estimate
the contribution from the quarks involved in the earlier steps of the 
evolution we use the specific properties of 
the KMR approach~[18] which enables us to discriminate between the various 
components of the quark distributions~[21, 22].

We would like to point out that, in contrast with the our previous investigation~[17], 
the present study is based on the off-shell matrix elements of underlying partonic subprocesses,
where the virtualities of both incoming gluons and quark are properly taken into account.
Numerically, we will investigate the 
total and differential cross sections of the
inclusive and associated  jet prompt photon photoproduction 
and perform a systematic comparison of our 
predictions with the available
H1 and ZEUS data~[2--5]. Our additional goal is to
study specific kinematical properties of the photon-jet system
which are strongly related to the intrinsic partonic $k_T$.

The outline of our paper is following. In Section~2 we 
recall shortly the basic formulas of the $k_T$-factorization approach with a brief 
review of calculation steps. In Section~3 we present the numerical results
of our calculations and a discussion. Section~4 contains our conclusions.

\section{Theoretical framework} 

\subsection{The subprocesses under consideration} \indent 

In $ep$ collisions at HERA prompt photons can be produced by one 
of three mechanisms: a direct production, a single resolved production 
and via parton-to-photon fragmentation processes~[23].
The direct contribution to the $\gamma p \to \gamma + X$ process
is the Compton scattering on the quark (antiquark)
$$
  \gamma (k_1) + q (k_2) \to \gamma (p_{\gamma}) + q (p'), \eqno(1)
$$

\noindent
where the particles four-momenta are given in parentheses. It gives 
the ${\cal O}(\alpha^2_{em})$ order contribution to the hadronic
cross section. Here $\alpha_{em}$ is Sommerfeld's fine structure constant.
The single resolved subprocesses are
$$
  q (k_1) + g (k_2) \to \gamma (p_{\gamma}) + q (p'), \eqno(2)
$$
$$
  g (k_1) + q (k_2) \to \gamma (p_{\gamma}) + q (p'), \eqno(3)
$$
$$
  q (k_1) + \bar q (k_2) \to \gamma (p_{\gamma}) + g (p'). \eqno(4)
$$

\noindent
Since the parton distributions in a photon at leading-order have a behavior 
proportional to $\alpha_{em} \ln \mu^2/\Lambda_{\rm QCD}^2 \sim \alpha_{em}/\alpha_s$,
these subprocesses give also
the ${\cal O}(\alpha^2_{em})$ contributions and 
therefore should be taken into account in our analysis. 

The calculation of the off-shell matrix elements (1) --- (4) is a very straightforward.
Here we would like to only mention two technical points. First,
in according to the $k_T$-factorization prescription~[11],
the summation over the incoming off-shell gluon polarizations is 
carried with $\sum \epsilon^\mu \epsilon^{  \, \nu} = {\mathbf k}_T^{\mu} {\mathbf k}_T^{\nu}/{\mathbf k}_T^2$,
where ${\mathbf k}_T$ is the gluon transverse momentum.
Second, when we calculate the matrix element squared,
the spin density matrix for all on-shell spinors is taken in the standard 
form $u (p) \bar u (p) = \hat p + m$.
In the case of off-shell initial quarks the on-shell spin density matrix has to
be replaced with a more complicated expression~[24]. To evaluate it,
we "extend" the original diagram and consider the off-shell quark line as internal line in
the extended diagram. The "extended" process looks like follows:
the initial on-shell quark with four-momentum $p$ and mass $m_q$ radiates a quantum (say,
photon or gluon) and becomes an off-shell quark with four-momentum $k$.
So, for the extended diagram squared we write:
$$
  |{\cal M}|^2 \sim {\rm Sp} \left[ {\bar {\cal T}^\mu} \, {\hat k + m_q \over k^2 - m_q^2}\,\gamma^\nu \, u(p) \bar u (p) \, \gamma_\nu {\hat k + m_q \over k^2 - m_q^2} \, {\cal T}_{\mu} \right], \eqno(5)
$$

\noindent
where ${\cal T}$ is the rest of the original matrix element which remains unchanged.
The expression presented between $\bar {\cal T}^\mu$ and ${\cal T}_\mu$ now plays the role
of the off-shell quark spin density matrix. Using the on-shell condition 
$u(p) \, \bar u(p) = \hat p + m_q$
and performing the Dirac algebra one obtains in the massless limit $m_q \to 0$:
$$
  |{\cal M}|^2 \sim {1\over (k^2)^2} {\bar {\cal T}^\mu} \, \left( 2 k^2 \hat p - 4 (p \cdot k) \hat k \right) \, {\cal T}_{\mu}. \eqno(6)
$$

\noindent
Now we use the Sudakov decomposition $k = x p + k_T$ and neglect the second term 
in the parentheses in (6) in the small-$x$ limit to arrive at
$$
  |{\cal M}|^2 \sim {2\over x k^2} {\bar {\cal T}^\mu} \, x \hat p \, {\cal T}_{ \,\mu}. \eqno(7)
$$

\noindent
(Essentially, we have neglected here the negative light-cone momentum fraction of the 
incoming quark). The properly normalized off-shell spin density matrix is given by $x \hat p$,
while the factor $2/xk^2$ has to be attributed to the
quark distribution function (determining its
leading behavior). With this normalization, we successfully recover the on-shell collinear
limit when $k$ is collinear with $p$.

\subsection{The CCFM and KMR unintegrated parton distributions} \indent 

As it was mentioned above, in the framework of $k_T$-factorization approach
one should consider the unintegrated gluon and quark distributions
$f_a(x,{\mathbf k}_T^2,\mu^2)$ instead of the conventional (collinear) 
parton densities $a(x,\mu^2)$.
In the KMR approximation, the unintegrated quark and 
gluon distributions are given by the expressions~[18]
$$
  \displaystyle f_q(x,{\mathbf k}_T^2,\mu^2) = T_q({\mathbf k}_T^2,\mu^2) {\alpha_s({\mathbf k}_T^2)\over 2\pi} \times \atop {
  \displaystyle \times \int\limits_x^1 dz \left[P_{qq}(z) {x\over z} q\left({x\over z},{\mathbf k}_T^2\right) \Theta\left(\Delta - z\right) + P_{qg}(z) {x\over z} g\left({x\over z},{\mathbf k}_T^2\right) \right],} \eqno (8)
$$
$$
  \displaystyle f_g(x,{\mathbf k}_T^2,\mu^2) = T_g({\mathbf k}_T^2,\mu^2) {\alpha_s({\mathbf k}_T^2)\over 2\pi} \times \atop {
  \displaystyle \times \int\limits_x^1 dz \left[\sum_q P_{gq}(z) {x\over z} q\left({x\over z},{\mathbf k}_T^2\right) + P_{gg}(z) {x\over z} g\left({x\over z},{\mathbf k}_T^2\right)\Theta\left(\Delta - z\right) \right],} \eqno (9)
$$

\noindent
where $P_{ab}(z)$ are the usual unregulated leading order DGLAP splitting 
functions, $q(x,\mu^2)$ and $g(x,\mu^2)$ are the conventional quark 
and gluon densities, $T_q({\mathbf k}_T^2,\mu^2)$ and $T_g({\mathbf k}_T^2,\mu^2)$ are 
the quark and gluon Sudakov form factors, and
the theta function $\Theta(\Delta - z)$ implies the angular-ordering constraint 
$\Delta = \mu/(\mu + |{\mathbf k}_T|)$ specifically to the last evolution step to 
regulate the soft gluon singularities~[18]. 

Another the solution for the unintegrated gluon distributions
have been obtained in~[25] from the CCFM evolution 
equation where all input parameters 
have been fitted to describe the proton structure function $F_2(x, Q^2)$.
The proposed gluon densities (namely, sets A0 and B0) have been applied to the 
number of QCD processes in the framework of the Monte-Carlo generator \textsc{cascade}~[20]
and in our calculations~[21].

In the present paper we will use both these distributions in our calculations.
To accomplish the CCFM-evolved gluon densities, one should apply the relevant 
unintegrated quark distributions. Below we will use the following 
way to get the $f_q(x,{\mathbf k}_T^2,\mu^2)$. 
The unintegrated valence quark densities $f_q^{(v)}(x,{\mathbf k}_T^2,\mu^2)$
have been obtained recently~[19] from the numerical solution of the CCFM-like 
equation\footnote{Authors are very grateful to Hannes Jung for providing us the code
for the unintegrated valence quark distributions.}. 
To calculate the contribution of the
sea quarks appearing at the last step of 
the gluon evolution, $f_q^{(g)}(x,{\mathbf k}_T^2,\mu^2)$,
we convolute the CCFM-evolved unintegrated gluon distribution 
$f_g(x,{\mathbf k}_T^2,\mu^2)$ with the standard leading-order DGLAP splitting function 
$P_{qg}(z)$:
$$
  f_q^{(g)}(x,{\mathbf k}_T^2,\mu^2) = {\alpha_s({\mathbf k}_T^2)\over 2 \pi} \int\limits_x^1 f_g(x/z,{\mathbf k}_T^2,\mu^2)P_{qg}(z)\,dz. \eqno(10)
$$

\noindent
Note that in the region of small ${\mathbf k}_T^2 < q_0^2$ the
scale in the strong coupling constant $\alpha_s$ is kept to be fixed at $q_0 = 1$~GeV.
To estimate the contribution of the sea quarks 
coming from the earlier evolution steps, $f_q^{(s)}(x,{\mathbf k}_T^2,\mu^2)$,
we apply the procedure based on the specific properties of the KMR 
scheme. Modifying~(8) in such a way that only the first term is 
kept and the second term is omitted and keeping only the sea quark in first term of~(8) we remove the 
valence and $f_q^{(g)}(x,{\mathbf k}_T^2,\mu^2)$ quarks from the evolution ladder.
In this way only the $f_q^{(s)}(x,{\mathbf k}_T^2,\mu^2)$ contribution to the $f_q(x,{\mathbf k}_T^2,\mu^2)$ 
is taken into account. 

We would like to point out that the valence quark densities from the CTEQ 6.1 set
have been used~[19] as the starting distributions to calculate the CCFM-evolved valence quark distributions
in a proton.
However, the CTEQ collaboration does not provide the quark and gluon distributions in a photon
(which are necessary to calculate the resolved photon contributions), and there is no
CCFM-evolved unintegrated quark densities in a photon. Therefore everywhere in our 
numerical analysis below we apply the KMR approximation for the unintegrated parton densities 
in a photon. Numerically, in~(8) and (9) we have tested the standard GRV-94 (LO)~[26] and MSTW-2008 (LO)~[27] sets
of collinear parton densities in a case of proton and the GRV-92 (LO)~[28] and CJKL (LO)~[29] sets in a case of photon.
To compare the different types of evolution, we have 
performed the numerical integration of the parton densities 
$f_a(x,{\mathbf k}_T^2,\mu^2)$ over transverse momenta ${\mathbf k}_T^2$.
In Fig.~1 we show the obtained "effective" valence quark 
distributions in a proton\footnote{The comparison of different unintegrated gluon densities 
to each other can be found, for example, in~[14--16].} as a function 
of $x$ for different values of $\mu^2$, namely 
$\mu^2 = 2\,{\rm GeV}^2$, $\mu^2 = 10\,{\rm GeV}^2$ and 
$\mu^2 = 100\,{\rm GeV}^2$.
The solid lines correspond to the CCFM-evolved unintegrated (valence) $u$-quark and $d$-quark densities. 
The dashed and dash-dotted lines correspond to the relevant KMR distributions
based on the collinear GRV-94 (LO) and MSTW-2008 (LO) sets, respectively.
We have observed some differences in both normalization and shape between the valence 
quark densities calculated within all these approaches.
Below we will study the dependence of our numerical results on the evolution scheme
in detail.

\subsection{Cross section for the prompt photon production} \indent 

Main formulas for prompt photon photoproduction have been obtained in our
previous paper~[17]. Here we only recall some of them.
Let $p_e$ and $p_p$ be the four-momenta of the initial electron and proton.
The direct contribution (1) to the $\gamma p \to \gamma + X$ 
process in the $k_T$-factorization approach can be written as
$$
  \sigma^{\rm (dir)} (\gamma p \to \gamma + X) = \sum_{q} \int {dx_2\over x_2} f_q(x_2,{\mathbf k}_{2 T}^2,\mu^2) d{\mathbf k}_{T}^2 {d\phi_2\over 2\pi} d{\hat \sigma} (\gamma q \to \gamma q), \eqno(11)
$$

\noindent 
where ${\hat \sigma} (\gamma q \to \gamma q)$ is the hard subprocess cross section 
via quark or antiquark having fraction $x_2$ of a initial proton longitudinal 
momentum, non-zero transverse momentum ${\mathbf k}_{2 T}$ 
(${\mathbf k}_{2 T}^2 = - k_{2 T}^2 \neq 0$) and azimuthal angle $\phi_2$.
The expression (11) can be easily rewritten in the form
$$
  \displaystyle \sigma^{\rm (dir)} (\gamma p \to \gamma + X) = \sum_q \int {E_T^\gamma \over 8\pi (x_2 s)^2 (1 - \alpha)} |\bar {\cal M}(\gamma q \to \gamma q)|^2 \times \atop 
  \displaystyle \times f_q(x_2,{\mathbf k}_{2 T}^2,\mu^2) dy^\gamma d E_T^\gamma d{\mathbf k}_{2 T}^2 {d\phi_2\over 2\pi} {d\phi^\gamma\over 2\pi}, \eqno(12)
$$

\noindent
where $|\bar {\cal M}(\gamma q \to \gamma q)|^2$ is the hard
matrix element squared which depends on the transverse momentum ${\mathbf k}_{2 T}^2$, 
$s = (k_1 + p_p)^2$ is the total energy of the subprocess under consideration, 
$y^\gamma$, $E_T^\gamma$ 
and $\phi^\gamma$ are the rapidity, transverse energy and azimuthal angle of the 
produced photon in the $\gamma p$ center-of-mass frame, and $\alpha = E_T^\gamma \exp y^\gamma /\sqrt s$. 

The formula for the 
resolved contribution to the prompt photon photoproduction in the 
$k_T$-factorization approach can be obtained by the similar way. But one should keep in mind that the
convolution in (11) should be made also with the unintegrated parton distributions 
$f_a^\gamma(x,{\mathbf k}_{T}^2,\mu^2)$ in a photon, i.e.  
$$
  \displaystyle d\sigma^{\rm (res)} (\gamma p \to \gamma + X) = \sum_{a,b} \int {dx_1\over x_1} f_a^\gamma(x_1,{\mathbf k}_{1 T}^2,\mu^2) d{\mathbf k}_{1 T}^2 {d\phi_1\over 2\pi} \times \atop 
  \displaystyle \times \int {dx_2\over x_2} f_b(x_2,{\mathbf k}_{2 T}^2,\mu^2) d{\mathbf k}_{2 T}^2 {d\phi_2\over 2\pi} d{\hat \sigma} (a b \to \gamma c), \eqno(13)
$$

\noindent
where $a, b, c = q$ and/or $g$, ${\hat \sigma} (a b \to \gamma c)$ is the cross section 
of the photon production in the corresponding parton-parton interaction (2) --- (4). 
Here parton $a$ has fraction $x_1$ of a initial photon longitudinal 
momentum, non-zero transverse momentum ${\mathbf k}_{1 T}$ 
(${\mathbf k}_{1 T}^2 = - k_{1 T}^2 \neq 0$) and azimuthal angle $\phi_1$.
We can easily obtain the final expression from equation (13). It has the form
$$
  \displaystyle \sigma^{\rm (res)} (\gamma p \to \gamma + X) = \sum_{a,b} \int {E_T^\gamma\over 8\pi (x_1 x_2 s)^2} |\bar {\cal M}(ab \to \gamma c)|^2 \times \atop 
  \displaystyle \times f_a^\gamma(x_1,{\mathbf k}_{1 T}^2,\mu^2) f_b(x_2,{\mathbf k}_{2 T}^2,\mu^2) d{\mathbf k}_{1 T}^2 d{\mathbf k}_{2 T}^2 dE_T^\gamma dy^\gamma dy^c {d\phi_1\over 2\pi} {d\phi_2\over 2\pi} {d\phi^\gamma\over 2\pi}, \eqno(14)
$$

\noindent
where $y^c$ is the rapidity of the parton $c$ in the $\gamma p$ center-of-mass frame. It is 
important that the hard matrix elements squared $|\bar {\cal M}(ab \to \gamma c)|^2$ depend
on the transverse momenta ${\mathbf k}_{1 T}^2$ and ${\mathbf k}_{2 T}^2$.
We would like to note that if we average the expressions (12) and (14) over 
$\phi_{1}$ and $\phi_{2}$ and take 
the limit ${\mathbf k}_{1 T}^2 \to 0$ and ${\mathbf k}_{2 T}^2 \to 0$,
then we obtain well-known expressions for 
the prompt photon production in leading-order (LO) perturbative QCD.

The experimental data~[2--5] taken by the H1 and ZEUS collaborations refer to  prompt
photon production in the $ep$ collisions, where the electron 
is scattered at small angle and the mediating photon is almost real ($Q^2 \sim 0$).
Therefore the $\gamma p$ cross sections (12) and (14)
need to be weighted with the photon flux in the electron:
$$
  d\sigma(e p\to e' + \gamma + X) = \int f_{\gamma/e}(y) d\sigma(\gamma p \to \gamma + X) dy, \eqno (15)
$$

\noindent
where $y$ is a fraction of the initial electron energy taken by the photon in the laboratory frame,
and we use the Weizacker-Williams approximation for the bremsstrahlung photon
distribution from an electron:
$$
  f_{\gamma/e}(y) = {\alpha_{em} \over 2\pi}\left({1 + (1 - y)^2\over y}\ln{Q^2_{\rm max}\over Q^2_{\rm min}} + 
  2m_e^2 y\left({1\over Q^2_{\rm max}} - {1\over Q^2_{\rm min}} \right)\right). \eqno (16)
$$

\noindent
Here $m_e$ is the electron mass, 
$Q^2_{\rm min} = m_e^2y^2/(1 - y)^2$ and $Q^2_{\rm max} = 1\,{\rm GeV}^2$, 
which is a typical value for the photoproduction measurements at HERA.

The multidimensional integration in (12), (14) and (15) has been performed
by means of the Monte Carlo technique, using the routine \textsc{vegas}~[30].
The full C$++$ code is available from the authors on 
request\footnote{lipatov@theory.sinp.msu.ru}.

\subsection{Fragmentation contributions and isolation} \indent 

In order to reduce the huge background
from the secondary photons produced by the decays of $\pi^0$, $\eta$ 
and $\omega$ mesons the isolation criterion is introduced in the experimental analyses.
This criterion is the following. A photon is isolated if the 
amount of hadronic transverse energy $E_T^{\rm had}$, deposited inside
a cone with aperture $R$ centered around the photon direction in the 
pseudo-rapidity and azimuthal angle plane, is smaller than
some value $E_T^{\rm max}$:
$$
  \displaystyle E_T^{\rm had} \le E_T^{\rm max},\atop
  \displaystyle (\eta - \eta^{\gamma})^2 + (\phi - \phi^{\gamma})^2 \le R^2. \eqno(17)
$$

\noindent
The both H1 and ZEUS collaborations take $R = 1$, $E_T^{\rm max} = \epsilon E_T^\gamma$ with 
$\epsilon = 0.1$ in the experiments~[2--5]. 
The isolation criteria not only reduces the background 
but also significantly reduces the fragmentation components.
It was shown~[7, 8] that after applying the isolation cut the contribution from the
fragmentation subprocesses is only about 5 or 6\% of the total prompt photon cross section.
Therefore in our further analysis we will neglect the small fragmentation 
contribution and consider only the direct and resolved production mechanisms.

\section{Numerical results} \indent

We now are in a position to present our numerical results. First we describe our
theoretical input and the kinematical conditions. After we fixed the unintegrated
parton distributions in a proton and in a photon, the cross sections (12) and (14) depend on
the energy scale $\mu$. As it often done~[7, 8] for prompt photon production, 
we choose the renormalization and factorization scales to be $\mu = \xi E_T^\gamma$. 
In order to estimate the scale uncertainties of our calculations
we will vary the parameter $\xi$ between 1/2 and 2 about the default value $\xi = 1$.
We use LO formula for the strong coupling constant $\alpha_s(\mu^2)$ 
with $n_f = 4$ massless quark flavours and $\Lambda_{\rm QCD} = 200$ MeV, 
such that $\alpha_s(M_Z^2) = 0.1232$.

\subsection{Inclusive prompt photon photoproduction} \indent 

Experimental data for the inclusive prompt photon production at HERA
come from both the ZEUS and H1 collaborations\footnote{Very recently the H1 collaboration has presented
the data~[6] which have been analysed in the $k_T$-factorization approach supplemented with the 
KMR partons.}. Two differential cross section are determined: first
as a function of the transverse energy $E_T^\gamma$, and second as
a function of pseudo-rapidity $\eta^\gamma$. The ZEUS data~[2] refer to the
kinematic region\footnote{Here and in the following all kinematic quantities
are given in the laboratory frame where positive OZ axis direction is given by the proton beam.}
defined by $E_T^\gamma > 5$ GeV and $-0.7 < \eta^\gamma < 0.9$ with
electron energy $E_e = 27.5$ GeV and proton energy $E_p = 820$ GeV.
The fraction $y$ of the electron energy trasferred to the photon is restricted
to the range $0.2 < y < 0.9$. Additionally the available ZEUS data for the 
prompt photon pseudo-rapidity distributions have been given also for three 
subdivisons of the $y$ range, namely $0.2 < y < 0.32$ ($134 < W < 170$ GeV), 
$0.32 < y < 0.5$ ($170 < W < 212$ GeV) and $0.5 < y < 0.9$ ($212 < W < 285$ GeV).
The more recent H1 data~[5] refer to the kinematic region
defined by $5 < E_T^\gamma < 10$ GeV, $ - 1 < \eta^\gamma < 0.9$ and $0.2 < y < 0.7$
with electron energy $E_e = 27.6$ GeV and proton energy $E_p = 920$ GeV.

The transverse energy and pseudo-rapidity distributions of the inclusive 
prompt photon production for different kinematical regions are shown
in Figs.~2 --- 4 in comparison with the available HERA data~[2, 5]. 
The solid histograms correspond to the results obtained
using the KMR approximation for the unintegrated quark and gluon densities in a proton and in a photon
(supplemented with the GRV-94 and GRV-92 parametrizations, respectively).
The dashed and dash-dotted histograms correspond 
to the results obtained with the CCFM-evolved unintegrated quark $f_q^{(v)}(x,{\mathbf k}_T^2,\mu^2)$,  
$f_q^{(g)}(x,{\mathbf k}_T^2,\mu^2)$ and gluon $f_g(x,{\mathbf k}_T^2,\mu^2)$ distributions
in a proton, as it was described in Section~2.2. 
These calculations are based on the CCFM set A0 and set B0 gluon densities.
The numerical predictions have been obtained by fixing both the
factorization and normalization scales at the default value $\mu = E_T^\gamma$.
One can see that the H1 and ZEUS data~[2, 5] can be reasonably well described
by using the KMR unintegrated parton densities. 
This is in a full agreement with our previous observations~[17].
Our predictions tend even to slightly overshoot the ZEUS data
at high values of $y$ variable and large photon pseudo-rapidity $\eta^\gamma$ (see Fig.~4).
Concerning the CCFM predictions, the results coming from the CCFM and KMR parton
densities are very similar to each other 
in the forward region, $\eta^\gamma > 0.4$. 
However, we find the some underestimation of the 
HERA data in the rear pseudo-rapidity $\eta^\gamma$ region.
One of the possible reasons of such disagreement can be connected with the 
contributions from the sea quarks involved in the 
earlier steps of the evolution cascade (below we will refer to these 
contributions as to "reduced sea" component).
Since the "reduced sea" is not taken into account in the CCFM evolution
we use the properties of the KMR approach to perform a rough  numerical estimation
of this contribution (see Table 1 and 2), as it was described above in Section~2.2.
We found that the "reduced sea" component gives approximately 
15\% contribution to the calculated cross sections. 
However, to avoid  double counting we do not sum the CCFM predictions and 
the estimated "reduced sea" contributions since part of them can be
already included into the CCFM results (via initial 
parton distributions which enter to the CCFM equation).

\begin{table}
\begin{center}
\begin{tabular}{|l|c|c|}
\hline
  & & \\
  uPDF (proton) + uPDF (photon) & $\sigma$ (H1 region)~[pb] & $\sigma$ (ZEUS region)~[pb] \\
  & & \\
\hline
  & & \\
  KMR (GRV-94) + KMR$^{\gamma}$(GRV-92) & $45.76^{+4.02}_{-4.48}$ & $47.19^{+4.86}_{-4.96}$ \\
  & & \\
  KMR (MSTW) + KMR$^{\gamma}$(CJKL) & $36.58^{+1.71}_{-2.38}$ & $37.97^{+2.13}_{-2.53}$ \\
  & & \\
  CCFM (set A0) + KMR$^{\gamma}$(GRV-92) & $38.03^{+3.97}_{-2.15}$ & $41.20^{+2.97}_{-1.07}$ \\
  & & \\
  CCFM (set A0) + KMR$^{\gamma}$(CJKL) & $33.48^{+4.41}_{-2.74}$ & $36.11^{+3.59}_{-1.99}$ \\
  & & \\
  CCFM (set B0) + KMR$^{\gamma}$(GRV-92) & $33.40^{+4.07}_{-2.21}$ & $36.12^{+3.14}_{-1.26}$ \\
  & & \\
  CCFM (set B0) + KMR$^{\gamma}$(CJKL) & $29.37^{+4.12}_{-2.63}$ & $31.88^{+3.42}_{-1.92}$ \\ 
  & & \\
\hline
  & & \\
  "reduced sea" & $7.56$ & $7.99$ \\ 
  & & \\
\hline
\end{tabular}
\end{center}
\caption{The total cross section of inclusive prompt photon photoproduction 
obtained in the kinematic range of the H1 and ZEUS experiments. The theoretical 
uncertainties in the predictions correspond to the usual scale variations, as it was described in the text.}
\end{table}

The total cross sections of the inclusive prompt photon production are listed in Table~1.
To study the dependence of our results on the evolution scheme we vary the unintegrated 
parton densities both in a proton and in a photon, as it was described in Section~2.2.
Additionally we study the effect of scale variations in the
calculated cross sections. We found that this effect is rather large: the relative difference
between results for $\mu = E_T^\gamma$ and results for $\mu = E_T^\gamma/2$ or
$\mu = 2 E_T^\gamma$ is about 10\%.
In the kinematic region of the ZEUS 
experiment our numerical predictions 
obtained with the KMR parton densities are rather close to the ones coming from
the usual (based on the collinear factorization of QCD) NLO 
calculations~[7, 8]. 

\subsection{Prompt photon photoproduction in association with jet} \indent 

 To calculate the semi-inclusive prompt photon production rates
we apply the procedure which has been used previously in~[17].
The produced photon is
accompanied by a number of partons radiated in the course of the parton evolution.
As it has been noted in~[31], on the average the parton 
transverse momentum decreases from the hard interaction
box towards the proton. As an approximation, we assume that the parton $k'$ 
emitted in the last evolution step compensates the whole transverse momentum
of the parton participating in the hard subprocess, i.e. ${\mathbf k'}_{T} \simeq - {\mathbf k}_{T}$.
All the other emitted partons are collected together in the proton remnant, which
is assumed to carry only a negligible transverse momentum compared to ${\mathbf k'}_{T}$.
This parton gives rise to a final hadron jet with $E_T^{\rm jet} = |{\mathbf k'}_{T}|$
in addition to the jet produced in the hard subprocess. From these hadron jets
we choose the one carrying the largest transverse energy, and then compute the cross section of prompt
photon with an associated jets.

The experimental data for this process were obtained by the H1 and ZEUS
collaborations. The H1 collaboration presented the 
cross sections~[5] measured differentially as a function of 
$E_T^\gamma$, $E_T^{\rm jet}$,
and the pseudo-rapidities $\eta^\gamma$ and $\eta^{\rm jet}$ in the 
kinematic region defined by $5 < E_T^\gamma < 10$ GeV, $E_T^{\rm jet} > 4.5$~GeV, 
$ - 1 < \eta^\gamma < 0.9$, $ - 1 < \eta^{\rm jet} < 2.3$ and $0.2 < y < 0.7$ with
electron energy $E_e = 27.6$ GeV and proton energy $E_p = 920$~GeV.
The more recent ZEUS data~[4] refer to the kinematic region
defined by $5 < E_T^\gamma < 16$~GeV, $6 < E_T^{\rm jet} < 17$~GeV, $ - 0.74 < \eta^\gamma < 1.1$,
$ - 1.6 < \eta^{\rm jet} < 2.4$ and $0.2 < y < 0.8$ with the same electron and proton energies.

The results of our calculations are shown in Figs.~5 --- 8 in 
comparison with the HERA data. 
One can see that the situation is very similar to the inclusive production case.
The dstributions measured by the H1 collaboration
are reasonably well reproduced by our calculations supplemented with the KMR unintegrated parton densities. 
However, there is some discrepancy between the predictions and the ZEUS data. 
It seems that the origin of
this disagreement is connected with the lowest bin in the $E_T^\gamma$ distribution, where 
our theoretical results are about 2 times below the ZEUS measurements (see Fig.~5, right panel).
In order to investigate it in more detail, we have repeted the calculations with an additional cut 
on the photon transverse energy, namely $E_T^\gamma > 7$~GeV (keeping the other cuts the same
as before). Our results compared to the ZEUS data are shown in Fig.~9. We found a perfect 
agreement between the theoretical predictions (based on the KMR parton densities) 
and the data after applying this additional cut (see also~[4]). 
Note that the KMR-based results agree with the H1 measurements~[5] in a whole $E_T^\gamma$ range.

Concerning the CCFM predictions, we found again that they are below the HERA data. 
In our opinion, it is connected with the missing "reduced sea" component
(which gives about 20\% contribution to the total $\gamma$ + jet cross section, see Table 2).
Note also that the shape of all predicted pseudo-rapidity $\eta^{\rm jet}$ distributions 
(based on the CCFM as well as on the KMR unintegrated parton densities) coincide 
with the ones calculated in the collinear NLO pQCD approximation~[7, 8]. As it was pointed
out~[5], the shape of this distribution is not reproduced well by the LO pQCD calculations.
This fact demonstrates that the main part of the collinear 
high-order corrections is already included at LO level in 
$k_T$-factorization formalism (see also~[14--16] for more information).

Now we turn to the total cross section of the prompt photon and associated jet photoproduction
at HERA.
Results of our calculations within the framework of the $k_T$-factorization 
approach compared to the ZEUS experimental data~[4] are listed in Table~2.
Similar to the inclusive photon production case, in these calculations 
we study the dependence of the predicted cross sections 
on the evolution scheme and the relative effects of scale variations.
The measured cross sections are described reasonably well using the $k_T$-factorization approach
and the KMR-constructed unintegrated parton densities.

\begin{table}
\begin{center}
\begin{tabular}{|l|c|c|}
\hline
  & & \\
  Source & $\sigma(\gamma + {\rm jet})$~[pb] (region I) & $\sigma(\gamma + {\rm jet})$~[pb] (region II)\\
  & & \\
\hline
  & & \\
  ZEUS measurement~[4] & $33.1 \pm 3.0~{\rm (stat.)}^{+4.6}_{-4.2}~{\rm (syst.)}$ & $13.8 \pm 1.2~{\rm (stat.)}^{+1.8}_{-1.6}~{\rm (syst.)}$ \\
  & & \\
  NLO QCD~[7] & $23.3^{+1.9}_{-1.7}$ & $14.9^{+1.3}_{-1.0}$ \\
  & & \\
  NLO QCD~[8] & $23.5^{+1.7}_{-1.6}$ & $13.4^{+1.1}_{-0.9}$ \\
  & & \\
  KMR (GRV-94) + KMR$^{\gamma}$(GRV-92) & $23.10^{+2.46}_{-2.19}$ & $14.88^{+1.37}_{-1.17}$ \\
  & & \\
  KMR (MSTW) + KMR$^{\gamma}$(CJKL) & $19.28^{+1.75}_{-0.89}$ & $12.9^{+0.38}_{-0.44}$ \\
  & & \\
  CCFM (set A0) + KMR$^{\gamma}$(GRV-92) & $17.13^{+1.22}_{-1.22}$ & $11.11^{+0.70}_{-0.49}$ \\
  & & \\
  CCFM (set A0) + KMR$^{\gamma}$(CJKL) & $15.29^{+0.68}_{-1.05}$ & $10.06^{+0.45}_{-0.46}$ \\
  & & \\
  CCFM (set B0) + KMR$^{\gamma}$(GRV-92) & $15.68^{+1.01}_{-0.68}$ & $10.26^{+0.55}_{-0.10}$ \\
  & & \\
  CCFM (set B0) + KMR$^{\gamma}$(CJKL) & $13.85^{+0.81}_{-0.82}$ & $9.10^{+0.58}_{-0.40}$ \\ 
  & & \\
\hline
  & & \\
  "reduced sea" & $4.49$ & $3.11$ \\ 
  & & \\
\hline
\end{tabular}
\end{center}
\caption{The total cross section of prompt photon and associated jet photoproduction 
obtained in the kinematic range $Q^2 < 1$~GeV$^2$, $5 < E_T^\gamma < 16$~GeV, $6 < E_T^{\rm jet} < 17$~GeV, $ - 0.74 < \eta^\gamma < 1.1$,
$ - 1.6 < \eta^{\rm jet} < 2.4$ and $0.2 < y < 0.8$ (region I). 
An additional cut $E_T^\gamma > 7$~GeV is applied in the region II.}
\end{table}

The most important variables for testing the structure of colliding 
proton and photon are the longitudinal fractional momenta of partons in these particles.
In order to reconstruct the momentum fractions of the 
initial partons from measured quantities the observables $x_\gamma^{\rm obs}$ and
$x_p^{\rm obs}$ are introduced in the ZEUS analysis~[3, 4]:
$$
  \displaystyle x_\gamma^{\rm obs} = {E_T^\gamma e^{ - \eta^\gamma} + E_T^{\rm jet} e^{ - \eta^{\rm jet}} \over 2 y E_e}, \quad 
  \displaystyle x_p^{\rm obs} = {E_T^\gamma e^{\eta^\gamma} + E_T^{\rm jet} e^{\eta^{\rm jet}} \over 2 E_p}. \eqno(17)
$$

\noindent
The $x_\gamma^{\rm obs}$ distribution is particularly sensitive to the
photon structure function. It is known that at large $x_\gamma^{\rm obs}$ 
region ($x_\gamma^{\rm obs} > 0.85$) the cross
section is dominated by the contribution of processes with direct initial photons, whereas
at $x_\gamma^{\rm obs} < 0.85$ the resolved photon contributions dominate~[4, 5].
Instead of using the $x_\gamma^{\rm obs}$ and $x_p^{\rm obs}$ variables, the H1 collaboration 
refers~[5] to $x_\gamma^{\rm LO}$ and $x_p^{\rm LO}$ observables given by
$$
  \displaystyle x_\gamma^{\rm LO} = {E_T^\gamma ( e^{ - \eta^\gamma} + e^{ - \eta^{\rm jet}} )\over 2 y E_e}, \quad 
  \displaystyle x_p^{\rm LO} = {E_T^\gamma ( e^{\eta^\gamma} + e^{\eta^{\rm jet}} )\over 2 E_p}. \eqno(18)
$$

\noindent
It was argued~[5] that these quantities make explicit use only of the photon energy, which is better measured
than the jet energy. Our predictions for all these observables compared to the H1 and ZEUS data~[4, 5] are 
shown in Figs.~10 and~11. We conclude again that KMR predictions reasonable agree 
with the HERA data for both direct and resolved production mechanisms. The 
sizeble contribition from the "reduced sea" quarks appears only for the direct production
and practically negligible for the resolved one.

Further understanding of the process dynamics and in particular of the high-order correction effects 
may be obtained from the transverse correlation between the produced prompt photon and the jet.
Specially the H1 and ZEUS collaborations have measured~[3--5] the distribution on the component of the
prompt photon's momentum perpendicular to the jet direction in the transverse plane, i.e.
$$
  p_\perp = |{\mathbf p}_{T}^\gamma \times {\mathbf p}_{T}^{\rm jet}|/|{\mathbf p}_{T}^{\rm jet}| = E_T^\gamma \sin \Delta \phi, \eqno(19)
$$

\noindent
where $\Delta \phi$ is the difference in azimuth between the photon and the accompanying jet. 
The ZEUS collaboration have measured~[3] also the distribution on the $\Delta \phi$ angle.
In the collinear leading-order approximation, these distributions must be simply
 delta functions $\delta(p_\perp)$ and $\delta(\phi - \pi)$, 
since the produced photon and the jet are back-to-back
in the transverse plane. Taking into account the non-vanishing initial parton
transverse momentum leads to the violation
of this back-to-back kinematics in the $k_T$-factorization approach.
The normalised $p_\perp$ and $\Delta \phi$ distributions compared to the H1 and ZEUS data~[3--5] 
are shown in Figs.~12 and 13 separately for the regions 
$x_\gamma^{\rm LO} < 0.85$ and $x_\gamma^{\rm LO} > 0.85$ (in the case of
ZEUS measurements for $x_\gamma^{\rm obs} > 0.9$ only). One can see that 
both the CCFM and KMR predictions are consistent with the data for all $p_\perp$ values at 
$x_\gamma^{\rm LO} > 0.85$ (or $x_\gamma^{\rm obs} > 0.9$) and tend to underestimate 
the data in the large $p_\perp$ region at $x_\gamma^{\rm LO} < 0.85$.
However, this underestimation is not significant and therefore we can conclude
that the CCFM-evolved parton densities reasonably well simulates the intrinsic 
partonic $k_T$.
The $k_T$-factorization predictions depicted in Fig.~11 are very similar 
to the ones~[7] obtained in the collinear factorization of QCD
at NLO level. The NLO calculations performed by another group~[8]
give a better description of the $p_\perp$ distributions at $x_\gamma^{\rm LO} < 0.85$ 
than the ones~[7] since in this kinematical region the cross section is 
dominated by ${\cal O}(\alpha_s)$ corrections to the processes with resolved photons, which are not 
included in the calculations~[7].

As a final point, we should mention that the corrections for hadronisation
and multiple interactions have been taken into account in the NLO analysis
of the available HERA data~[2--5] performed in the framework of collinear factorization of QCD. 
The correction factors are typically 0.8 --- 1.2 depending on a bin.
These corrections are not taken into account in our consideration.

\section{Conclusions} \indent 

In the present paper 
the evaluated CCFM and KMR unintegrated quark and gluon densities
have been applied to the analysis of the recent experimental data on the prompt photon
photoproduction taken by the H1 and ZEUS collaborations at HERA.
Our consideration is based on the off-shell matrix elements of underlying partonic
subprocesses (where the transverse momenta of both quarks and gluons are properly 
taken into account) and  covers both inclusive and associated with the hadronic jet production 
rates. We have studied the dependences of our numerical results on the evolution scheme
and on the standard scale variations.
To evaluate the unintegrated quark densities within the CCFM dynamics
we have calculated separately the contribution of valence quarks, sea quarks appearing at the 
last step of the gluon evolution and sea quarks coming from the earlier gluon splittings. 
In first time the contribution from the last gluon splitting has been calculated
as a convolution of the CCFM-evolved unintegrated gluon distribution 
with the standard leading-order DGLAP splitting function $P_{qg}(z)$. The
contribution from the sea quarks involved into the earlier evolution steps
has been estimated in the framework of the KMR approximation.

We have found a reasonable agreement between our predictions and the available data.
The contributions to the total photon cross section from the quarks emerging from the earlier steps of the parton 
evolution rather than from the last gluon splitting are estimated
to be of 15 -- 20\% approximately. Additionally we have studied the specific kinematical 
properties of the photon-jet system which are strongly sensitive to the 
transverse momentum of incoming partons. We have demonstrated that
the $k_T$-factorization approach supplemented with the CCFM and KMR parton dynamics
reasonably well simulates the intrinsic partonic $k_T$.

Note that in our analysis we neglect the contribution from the fragmentation 
processes and from the direct box diagram ($\gamma g \to \gamma g$). 
As it was claimed in~[7], the direct box diagram, which is
formally of the next-to-next-to-leading order, gives approximately 6\% contribution 
to the total NLO cross section. The problem of taking into
account the contribution of box diagrams with initial off-shell gluons 
in the framework of $k_T$-factorization is still open.

\section*{Acknowledgements} \indent 

We thank S.P.~Baranov for his encouraging interest and useful discussions
and H.~Jung for providing us the CCFM code for unintegrated valence quark and gluon distributions, reading of the manuscript and very useful discussions.
The authors are very grateful to 
DESY Directorate for the support in the 
framework of Moscow --- DESY project on Monte-Carlo
implementation for HERA --- LHC.
A.V.L. was supported in part by the grants of the president of 
Russian Federation (MK-438.2008.2) and Helmholtz --- Russia
Joint Research Group.
Also this research was supported by the 
FASI of Russian Federation (grant NS-1456.2008.2),
FASI state contract 02.740.11.0244 and RFBR grant 08-02-00896-a.

\newpage

\begin{figure}
\begin{center}
\epsfig{figure=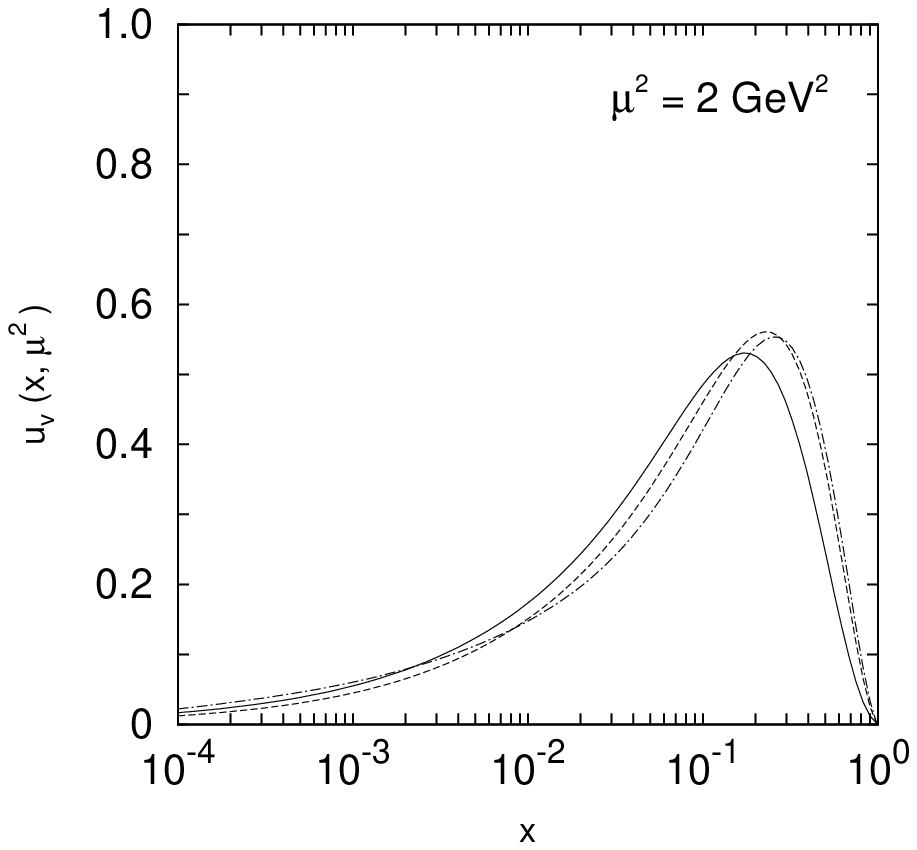, width = 8.1cm}
\epsfig{figure=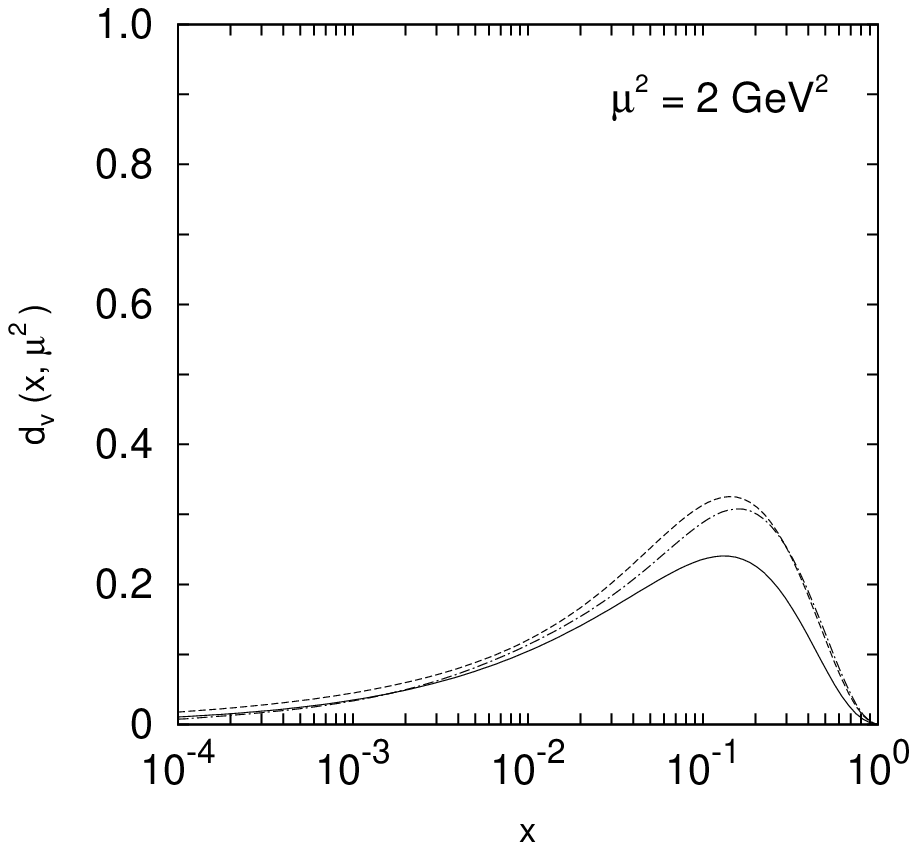, width = 8.1cm}
\epsfig{figure=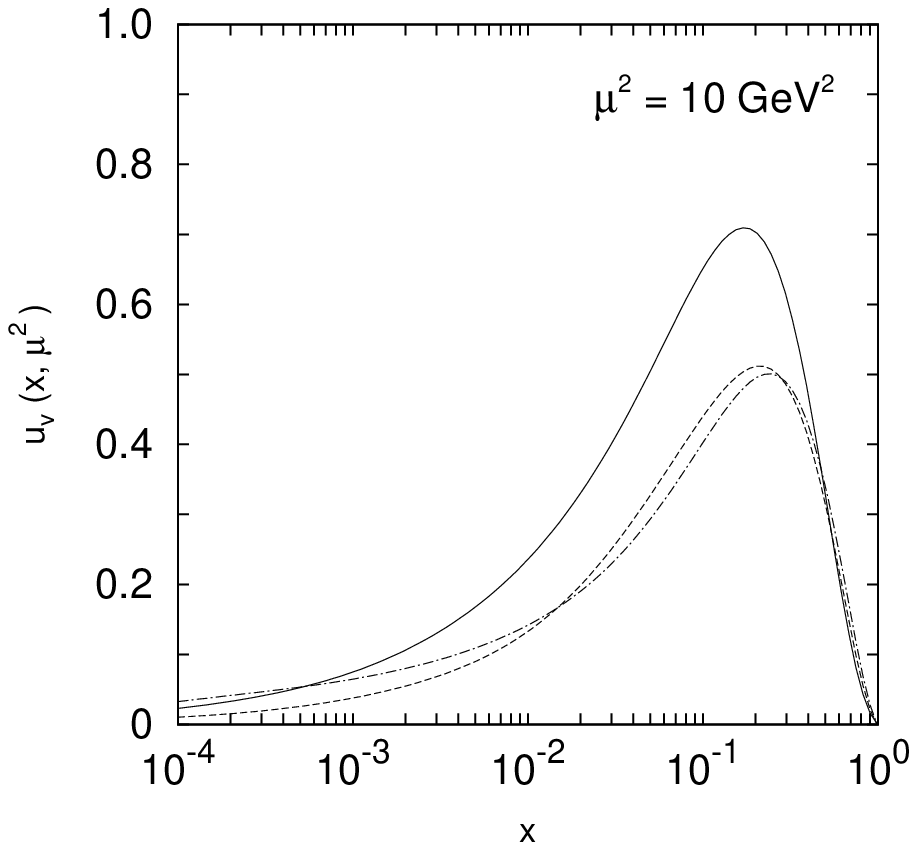, width = 8.1cm}
\epsfig{figure=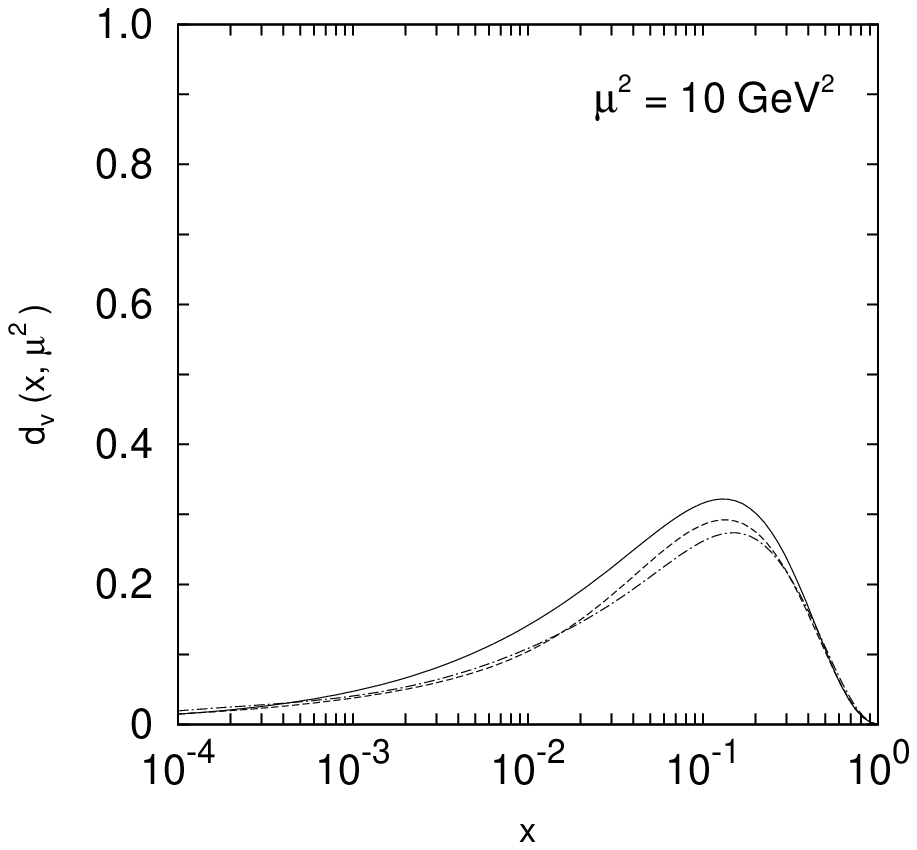, width = 8.1cm}
\epsfig{figure=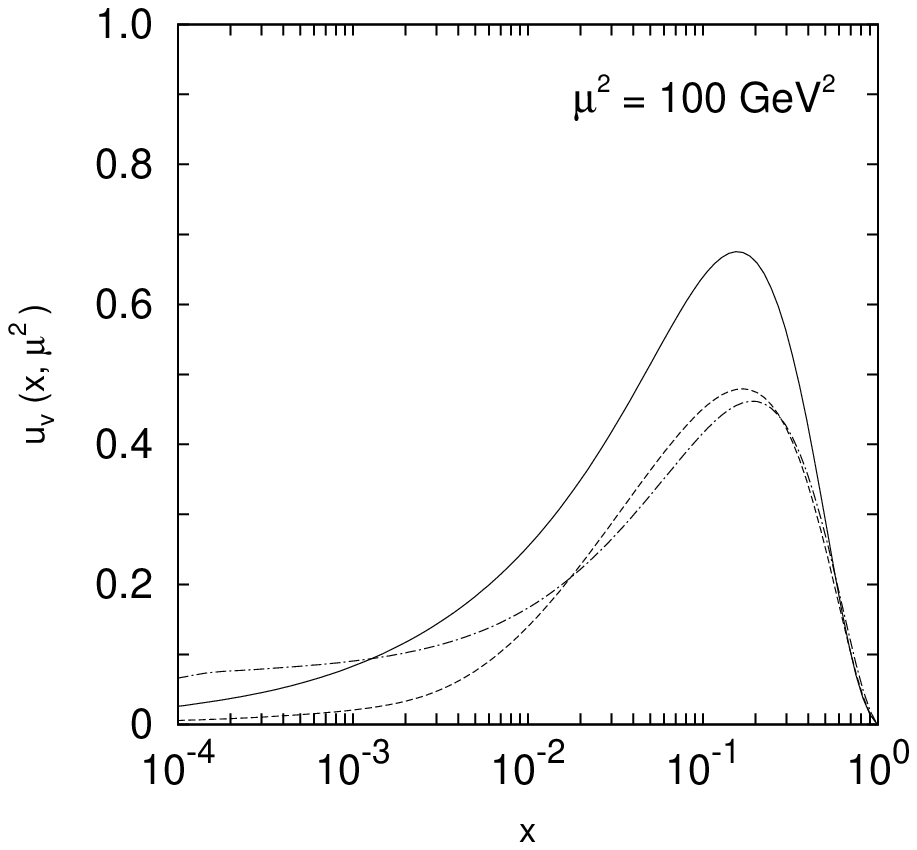, width = 8.1cm}
\epsfig{figure=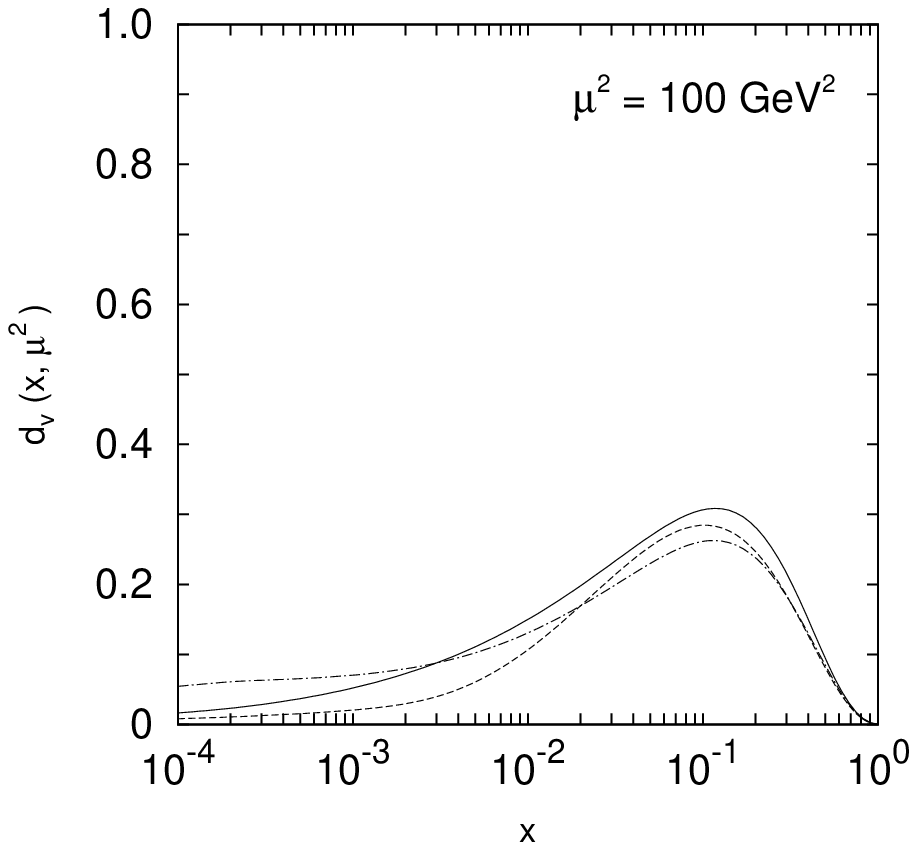, width = 8.1cm}
\caption{The effective (${\mathbf k}_T^2$-integrated) valence quark distributions in a proton
as a function of $x$ for different values of $\mu^2$.
The solid lines correspond to the CCFM-evolved quark distributions. 
The dashed and dash-dotted lines correspond to the KMR predictions
based on the collinear GRV-94 (LO) and MSTW-2008 (LO) sets, respectively.}
\end{center}
\label{fig0}
\end{figure}

\begin{figure}
\begin{center}
\epsfig{figure=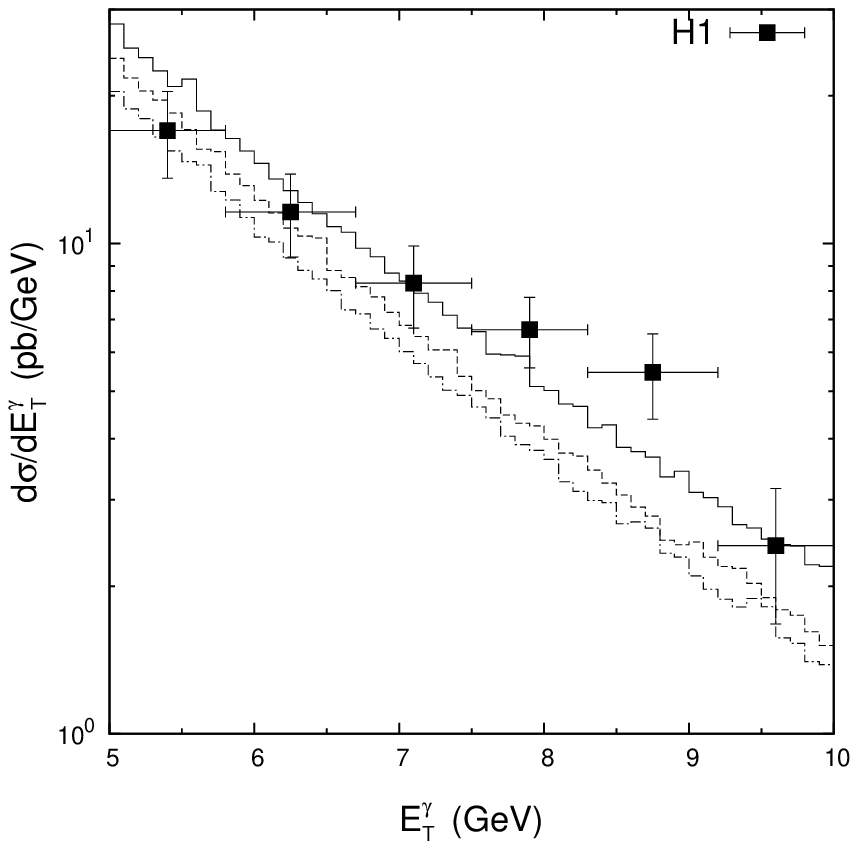, width = 8.1cm}
\epsfig{figure=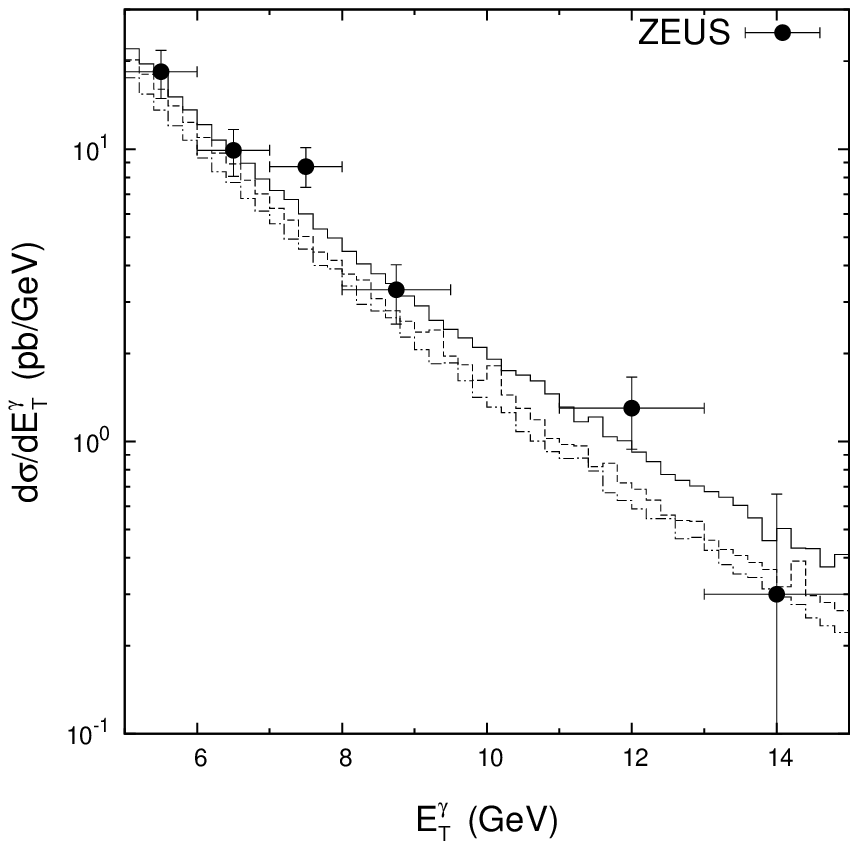, width = 8.1cm}
\caption{The differential cross sections $d\sigma/d E_T^\gamma$ for the 
inclusive prompt photon photoproduction at HERA.
The solid histograms correspond to the results obtained
using the KMR quark and gluon densities in a proton and in a photon.
The dashed and dash-dotted histograms correspond 
to the results obtained with the CCFM-evolved quark $f_q^{(v)}(x,{\mathbf k}_T^2,\mu^2)$,  
$f_q^{(g)}(x,{\mathbf k}_T^2,\mu^2)$ and gluon $f_g(x,{\mathbf k}_T^2,\mu^2)$ distributions
in a proton. In these calculations we use CCFM set A0 and set B0 gluons, respectively.
The experimental data are from H1~[5] and ZEUS~[2].}
\end{center}
\label{fig1}
\end{figure}

\begin{figure}
\begin{center}
\epsfig{figure=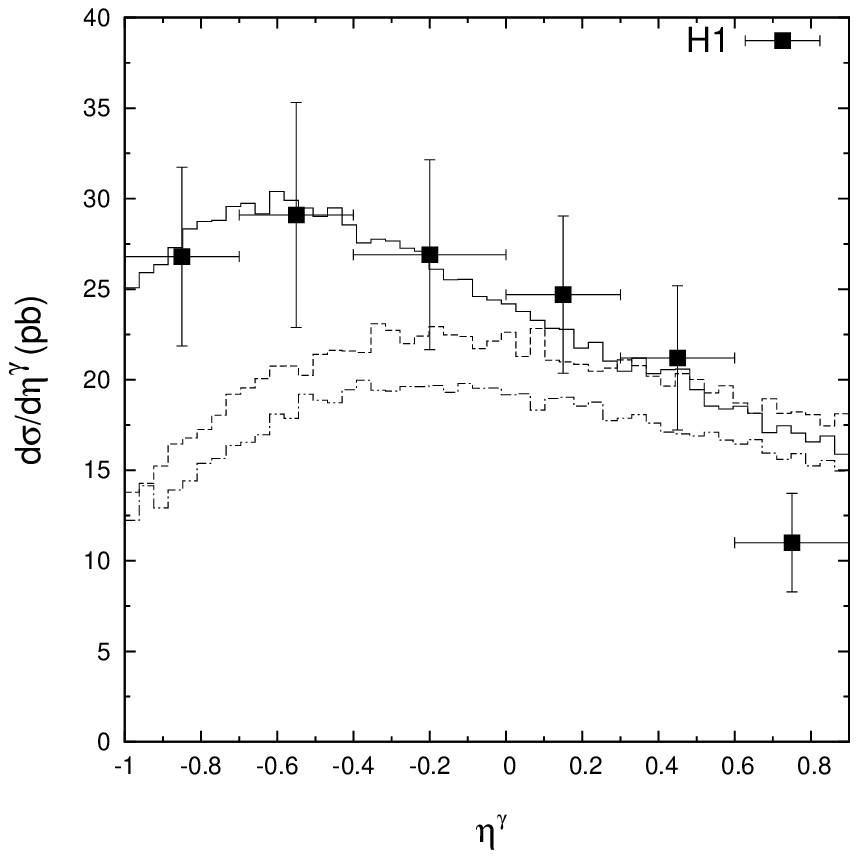, width = 8.1cm}
\epsfig{figure=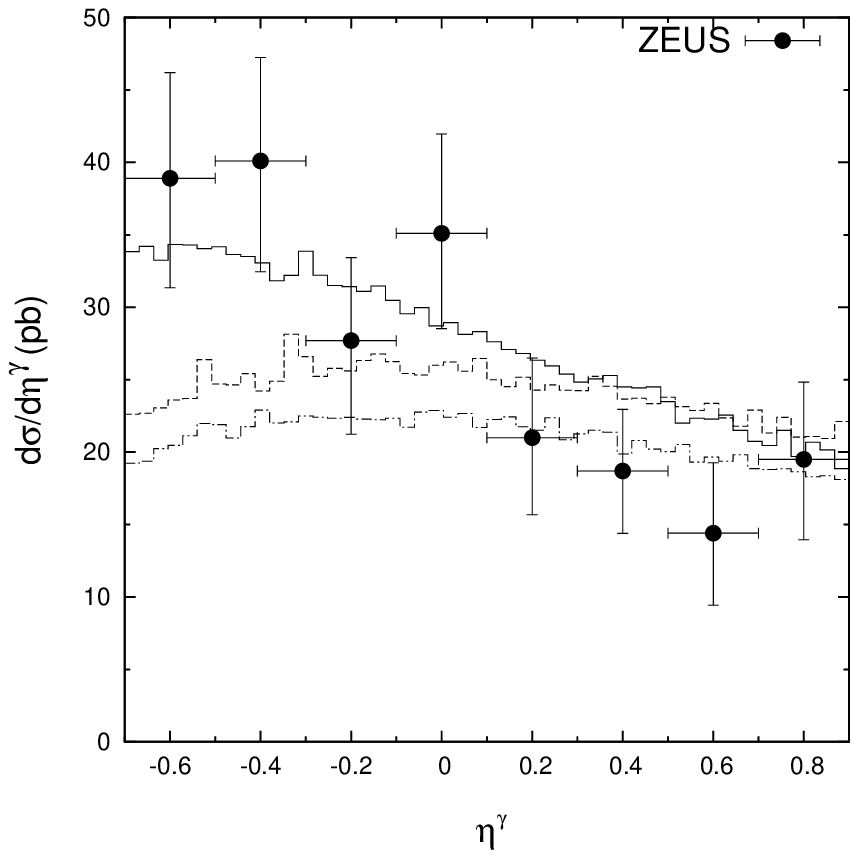, width = 8.1cm}
\caption{The differential cross sections $d\sigma/d\eta^\gamma$ for the 
inclusive prompt photon photoproduction at HERA. Notation of all histograms 
is the same as in Figure~2. 
The experimental data are from H1~[5] and ZEUS~[2].}
\end{center}
\label{fig2}
\end{figure}

\begin{figure}
\begin{center}
\epsfig{figure=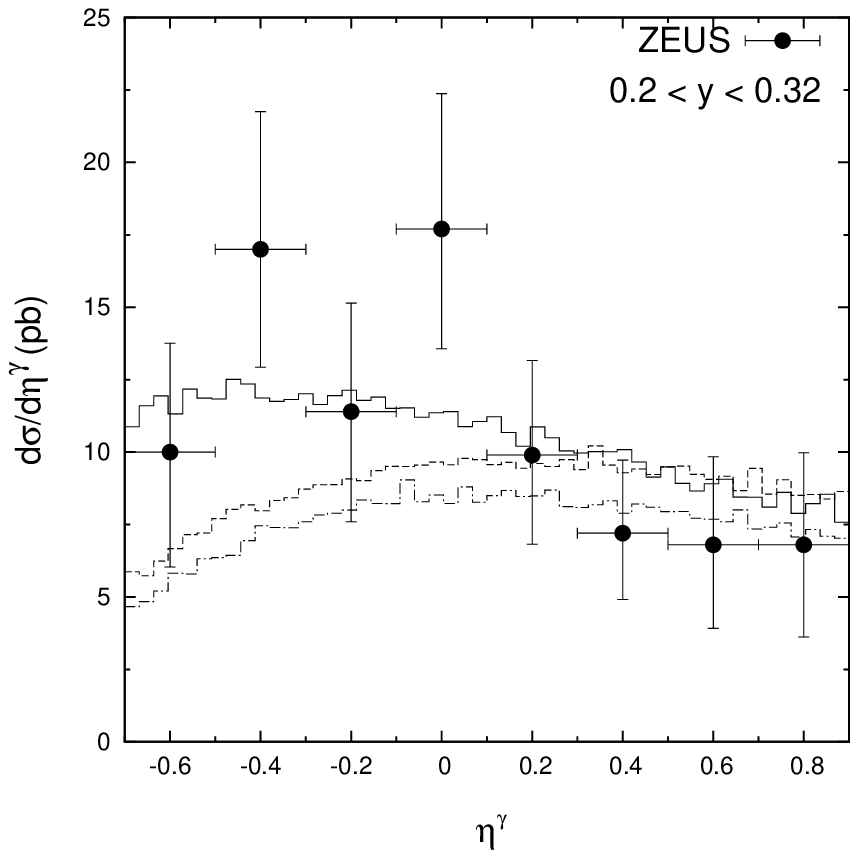, width = 8.1cm}
\epsfig{figure=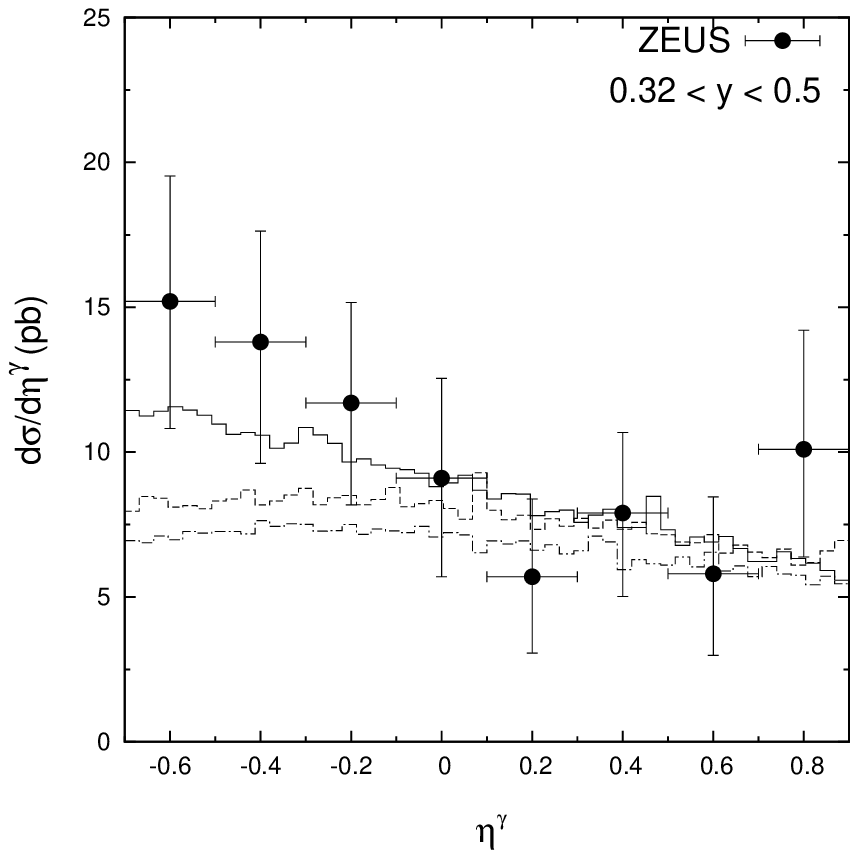, width = 8.1cm}
\epsfig{figure=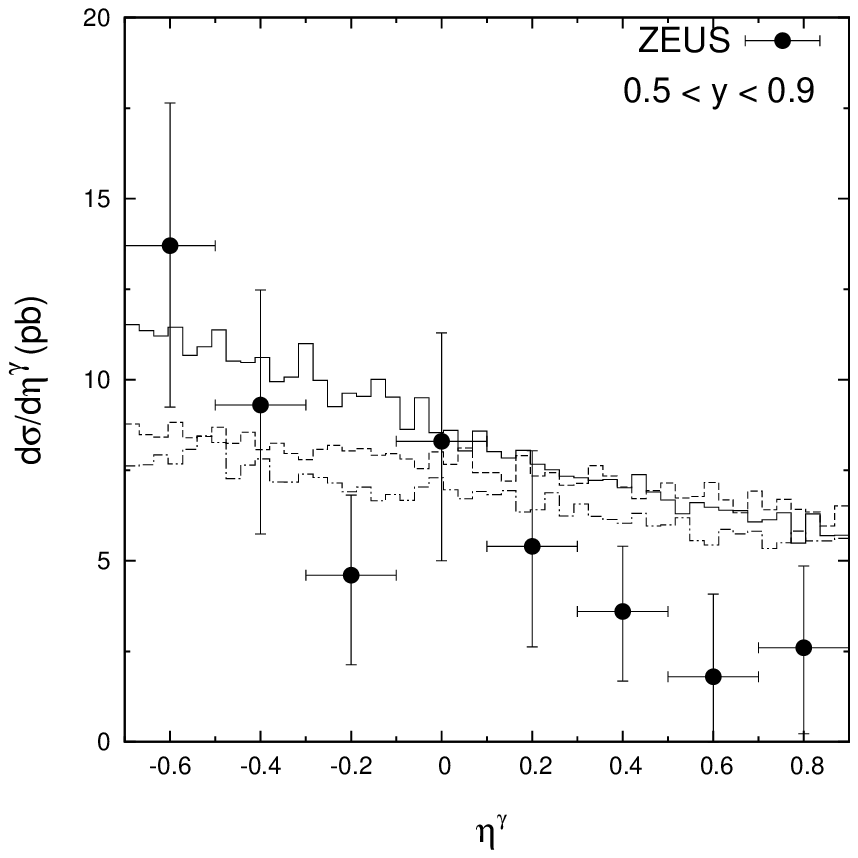, width = 8.1cm}
\caption{The differential cross sections $d\sigma/d\eta^\gamma$ for the 
inclusive prompt photon photoproduction at HERA calculated in the different kinematical regions. 
Notation of all histograms 
is the same as in Figure~2. 
The experimental data are ZEUS~[2].}
\end{center}
\label{fig3}
\end{figure}

\begin{figure}
\begin{center}
\epsfig{figure=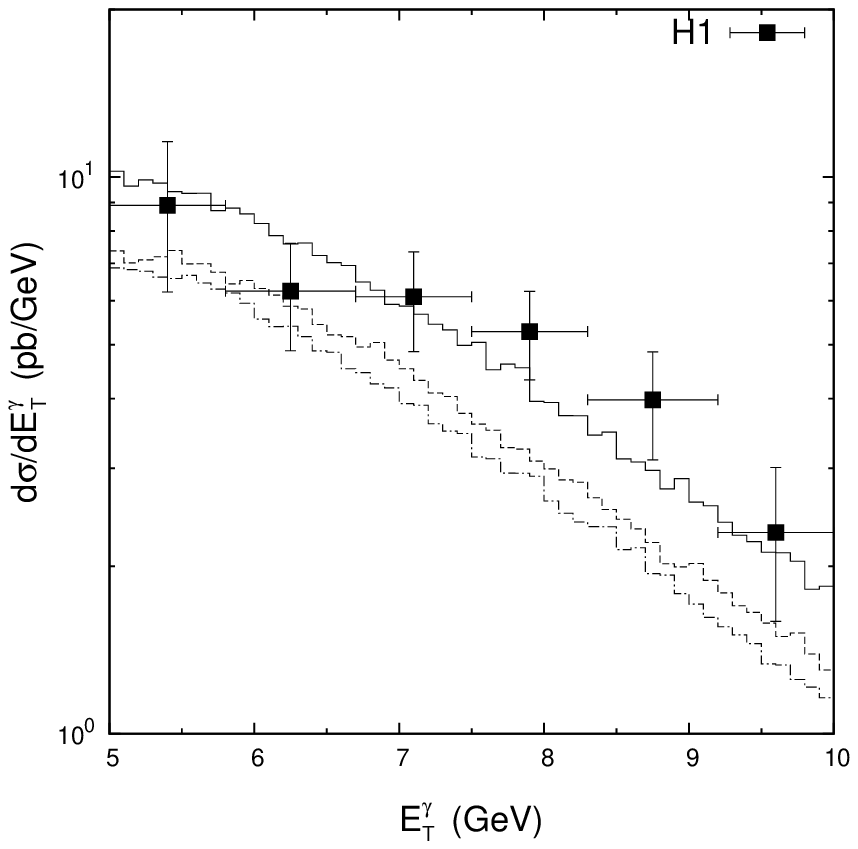, width = 8.1cm}
\epsfig{figure=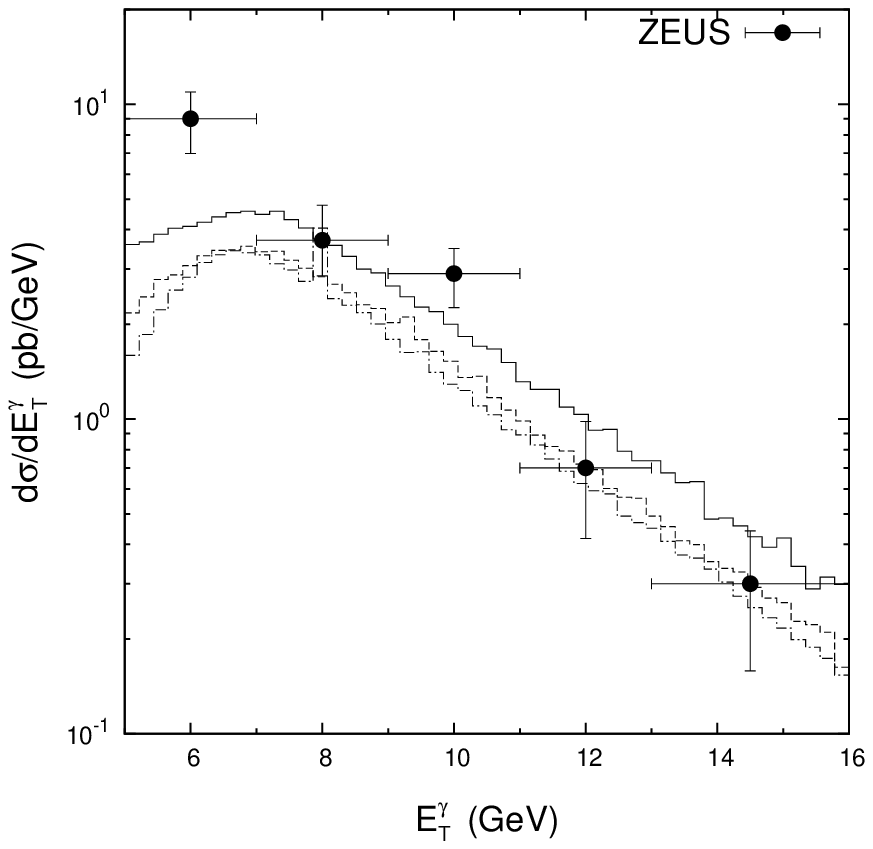, width = 8.1cm}
\caption{The differential cross sections $d\sigma/d E_T^\gamma$ for the 
prompt photon + jet production at HERA. Notation of all histograms 
is the same as in Figure~2. The experimental data are from H1~[5] and ZEUS~[4].}
\end{center}
\label{fig4}
\end{figure}

\begin{figure}
\begin{center}
\epsfig{figure=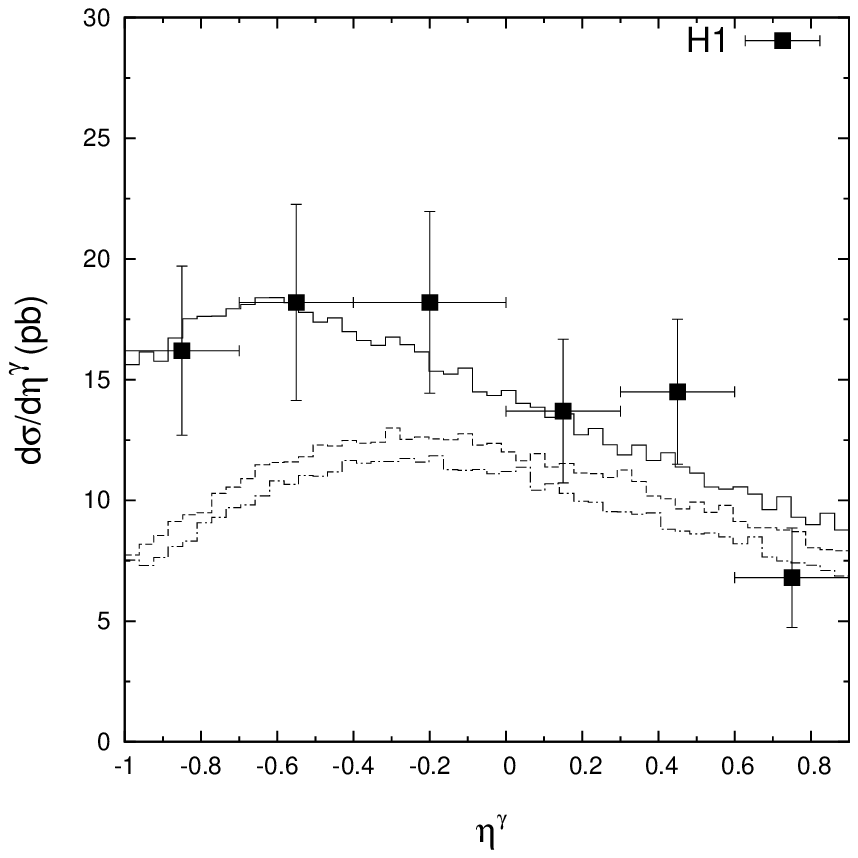, width = 8.1cm}
\epsfig{figure=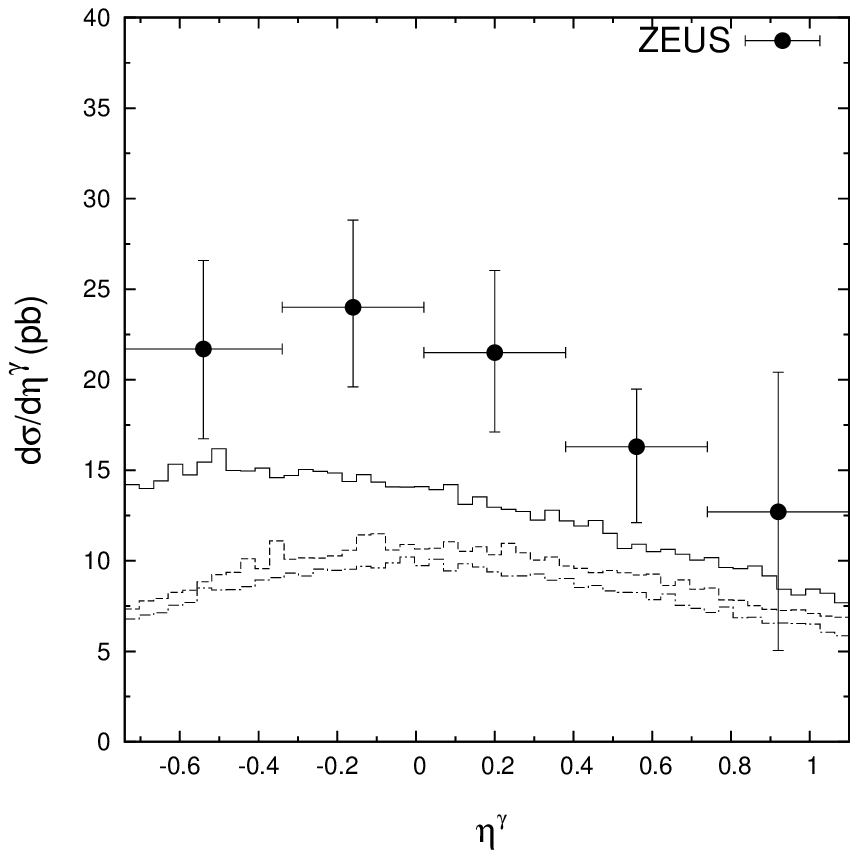, width = 8.1cm}
\caption{The differential cross sections $d\sigma/d\eta^\gamma$ for the 
prompt photon + jet production at HERA. Notation of all histograms 
is the same as in Figure~2. The experimental data are from H1~[5] and ZEUS~[4].}
\end{center}
\label{fig5}
\end{figure}

\begin{figure}
\begin{center}
\epsfig{figure=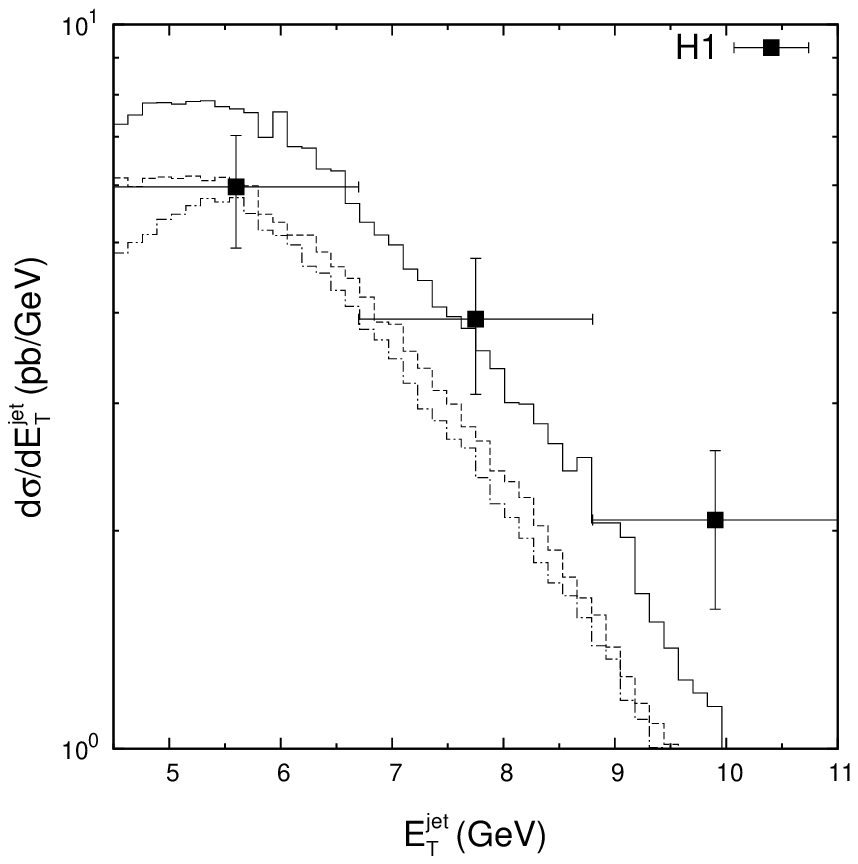, width = 8.1cm}
\epsfig{figure=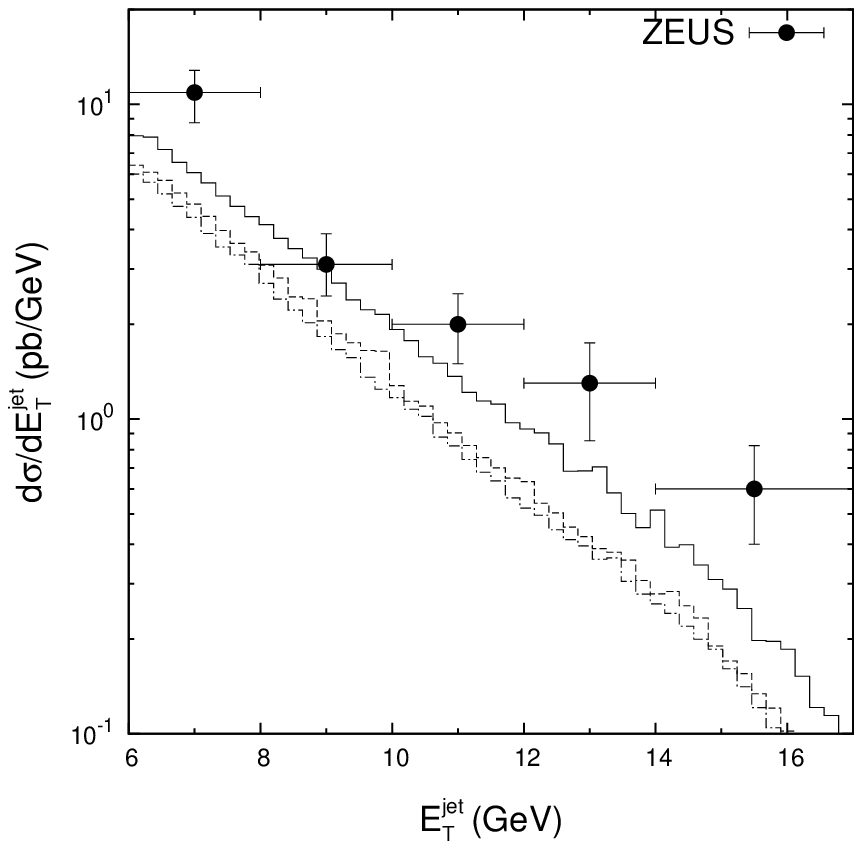, width = 8.1cm}
\caption{The differential cross sections $d\sigma/d E_T^{\rm jet}$ for the 
prompt photon + jet production at HERA. Notation of all histograms 
is the same as in Figure~2. The experimental data are from H1~[5] and ZEUS~[4].}
\end{center}
\label{fig6}
\end{figure}

\begin{figure}
\begin{center}
\epsfig{figure=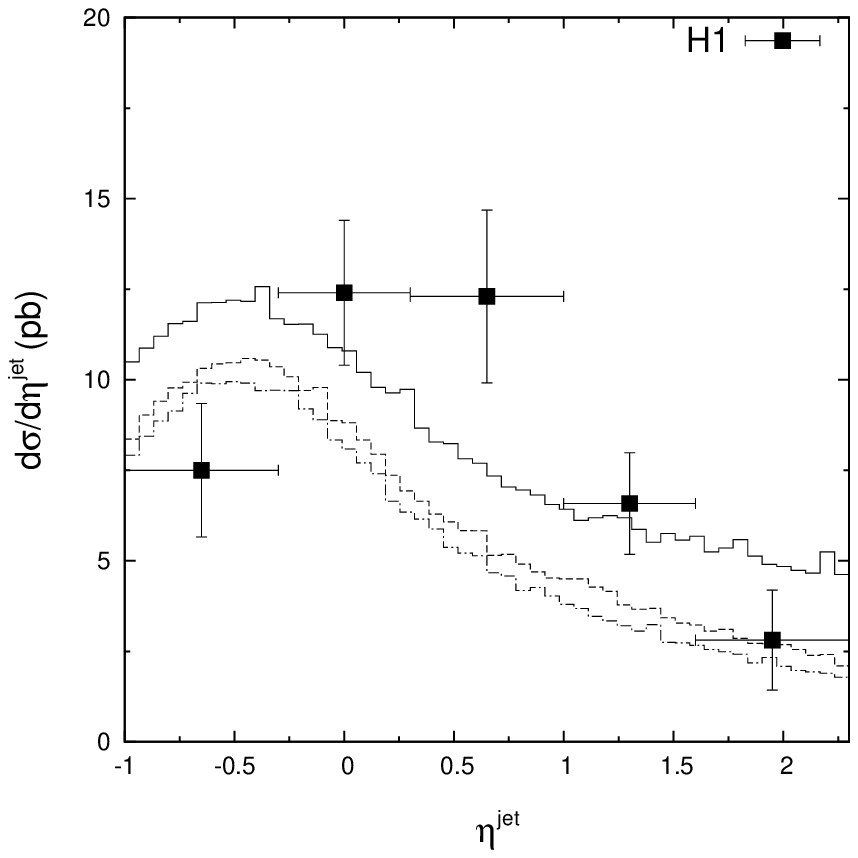, width = 8.1cm}
\epsfig{figure=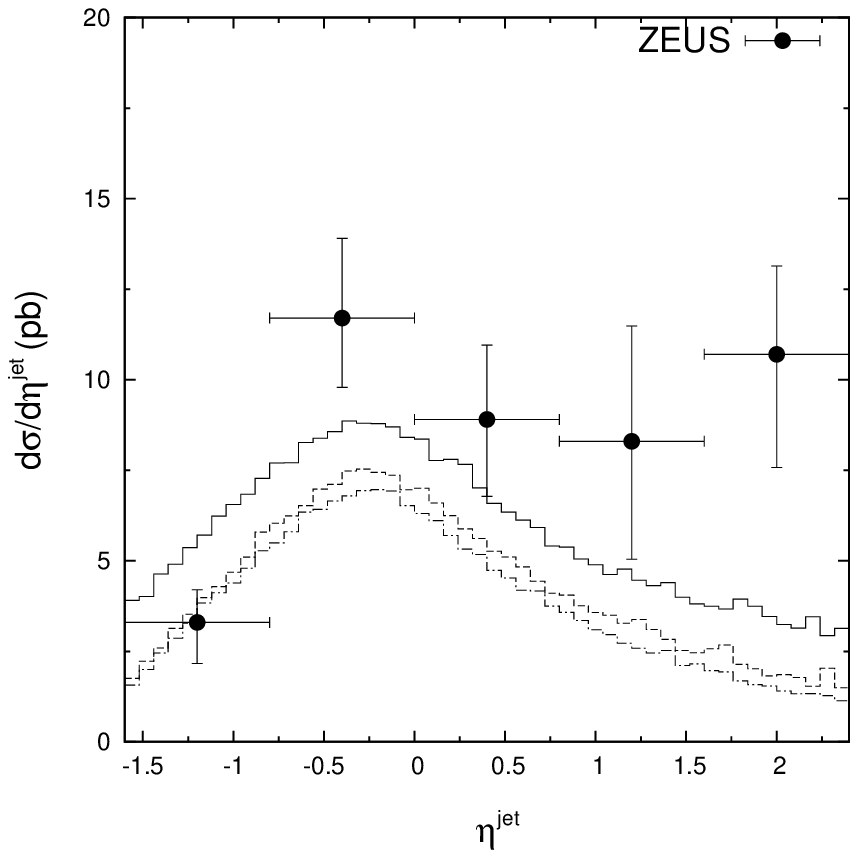, width = 8.1cm}
\caption{The differential cross sections $d\sigma/d \eta^{\rm jet}$ for the 
prompt photon + jet production at HERA. Notation of all histograms 
is the same as in Figure~2. The experimental data are from H1~[5] and ZEUS~[4].}
\end{center}
\label{fig7}
\end{figure}

\begin{figure}
\begin{center}
\epsfig{figure=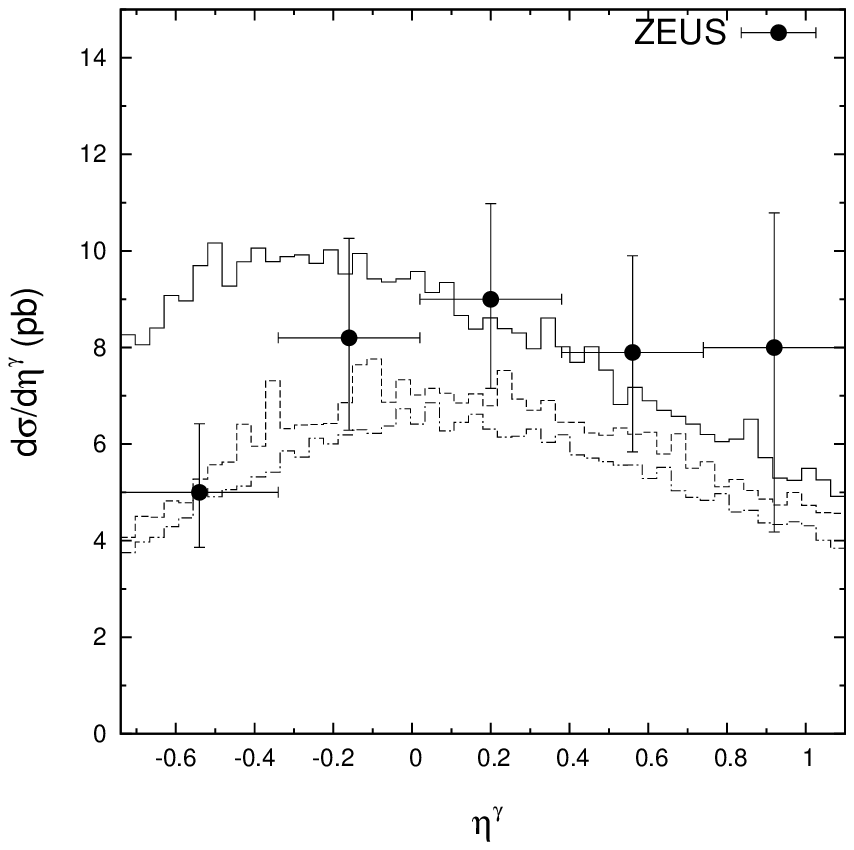, width = 8.1cm}
\epsfig{figure=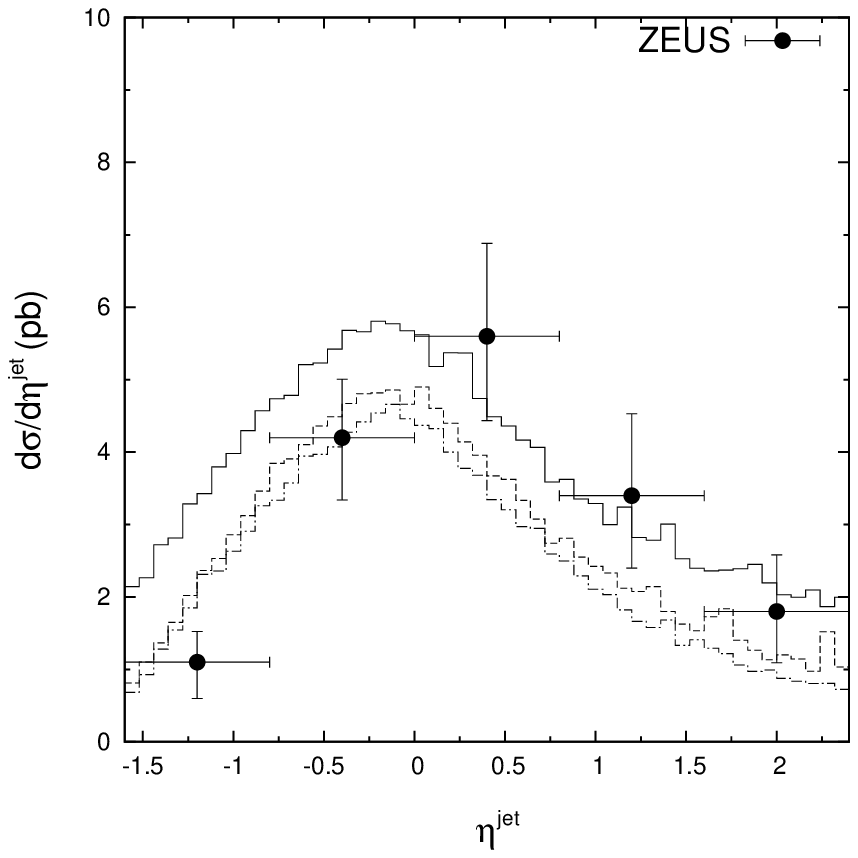, width = 8.1cm}
\epsfig{figure=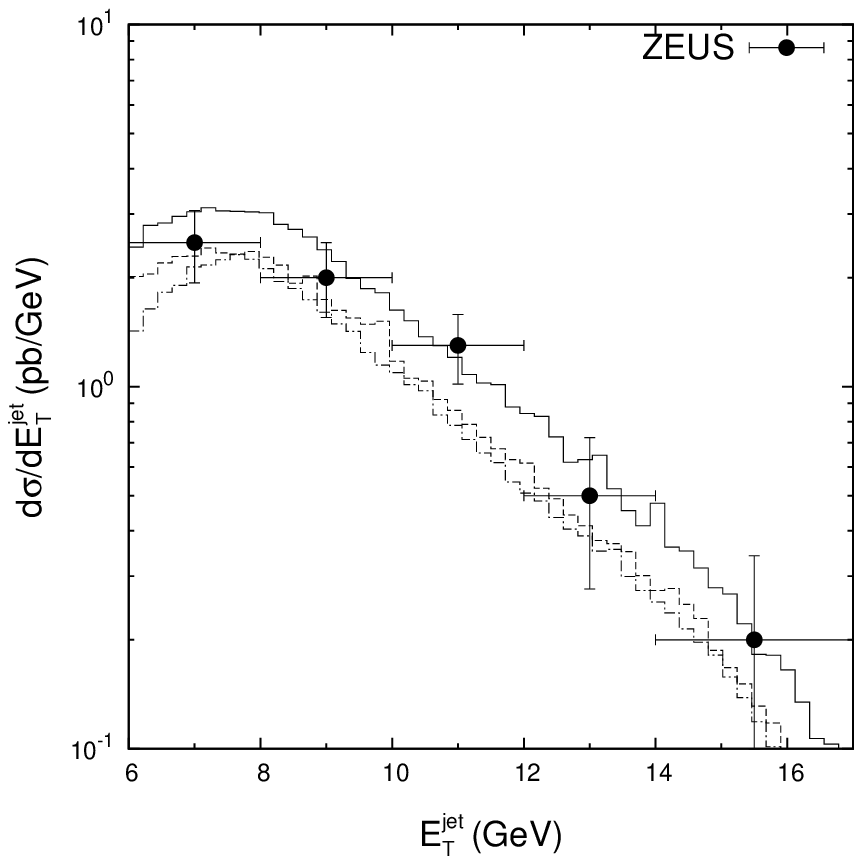, width = 8.1cm}
\caption{The differential cross sections $d\sigma/d \eta^\gamma$, $d\sigma/d \eta^{\rm jet}$ and 
$d\sigma/d E_T^{\rm jet}$ for the prompt photon + jet production at HERA. The additional cut 
$E_T^\gamma > 7$~GeV has been applied. Notation of all histograms 
is the same as in Figure~2. The experimental data are from ZEUS~[4].}
\end{center}
\label{fig8}
\end{figure}

\begin{figure}
\begin{center}
\epsfig{figure=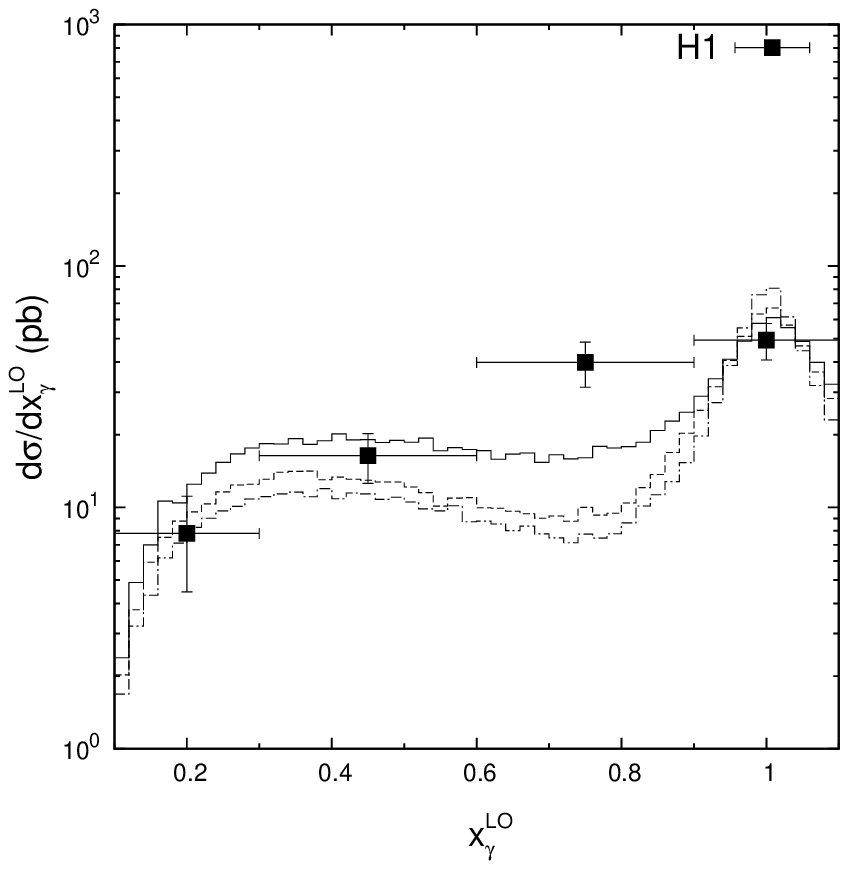, width = 8.1cm}
\epsfig{figure=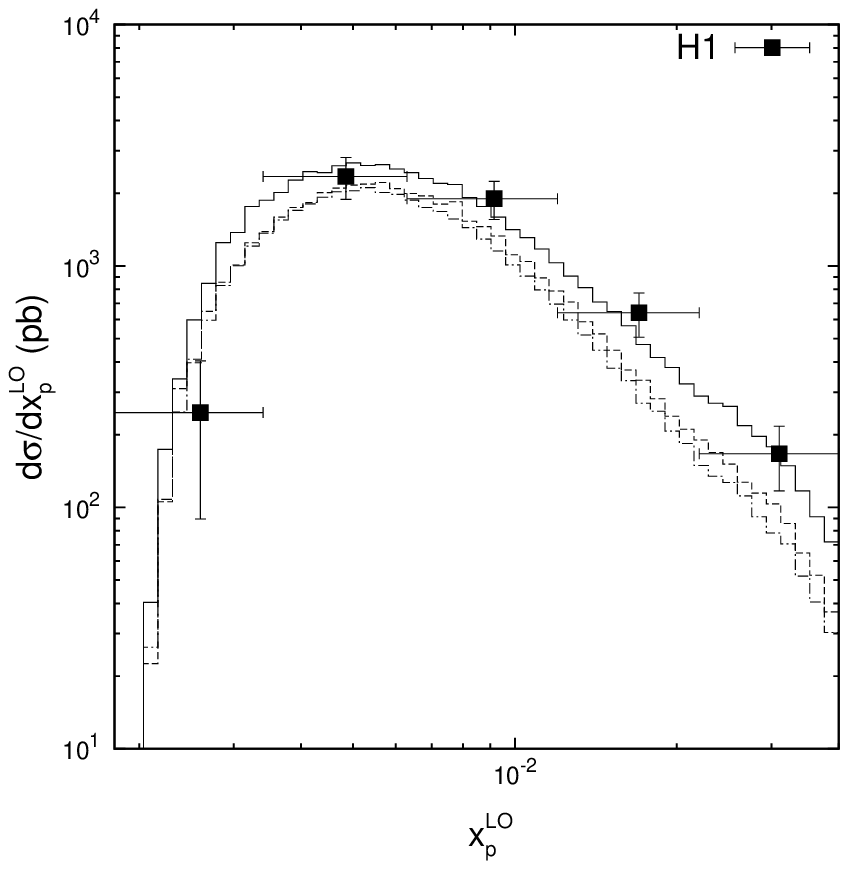, width = 8.1cm}
\caption{The differential cross sections $d\sigma/dx_\gamma^{\rm LO}$ and $d\sigma/dx_p^{\rm LO}$
for the prompt photon + jet production at HERA. Notation of all histograms 
is the same as in Figure~2. The experimental data are from H1~[5].}
\end{center}
\label{fig9}
\end{figure}

\begin{figure}
\begin{center}
\epsfig{figure=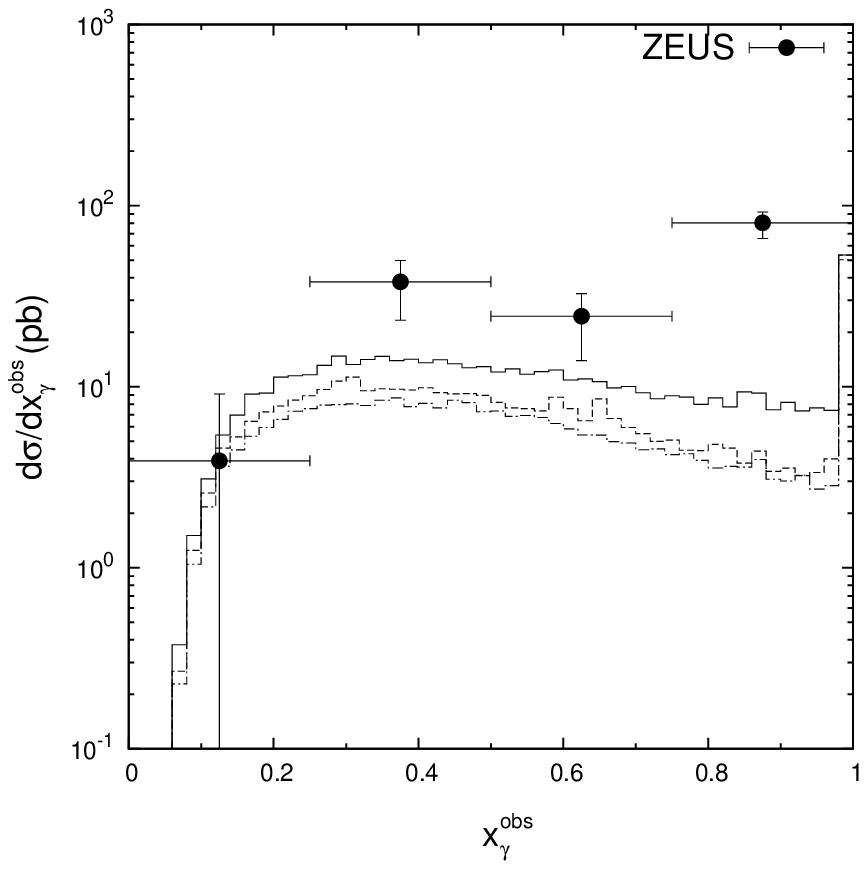, width = 8.1cm}
\epsfig{figure=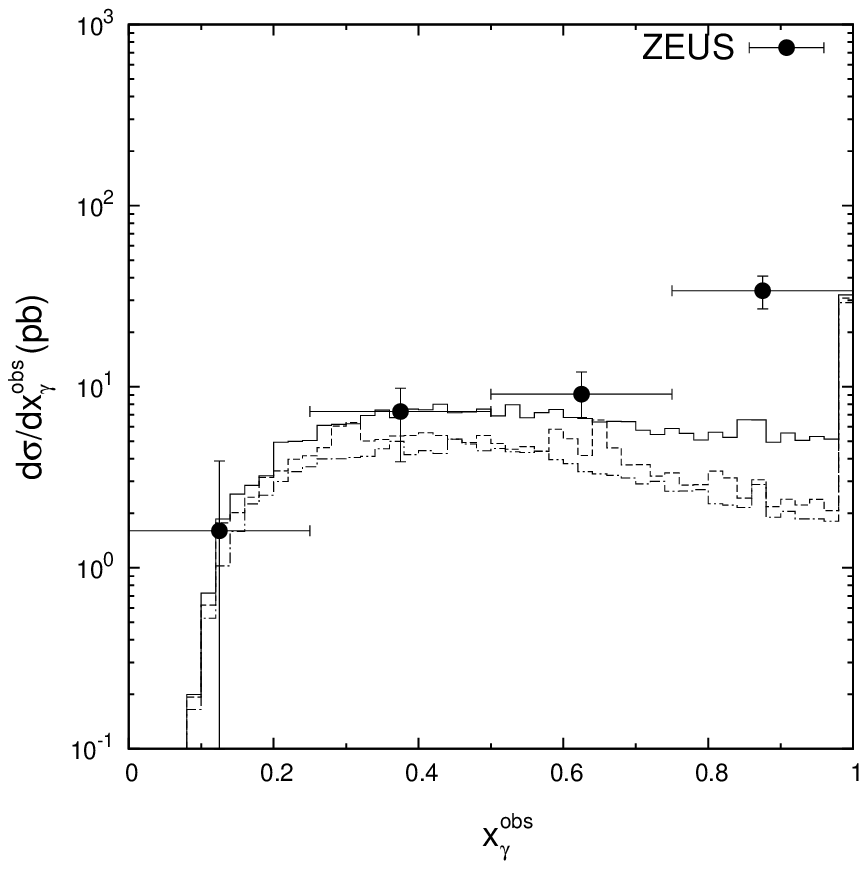, width = 8.1cm}
\caption{The differential cross sections $d\sigma/dx_\gamma^{\rm obs}$ 
for the prompt photon + jet production at HERA. Notation of all histograms 
is the same as in Figure~2. The additional cut $E_T^\gamma > 7$~GeV has been applied on the right panel.
The experimental data are from ZEUS~[4].}
\end{center}
\label{fig9a}
\end{figure}

\begin{figure}
\begin{center}
\epsfig{figure=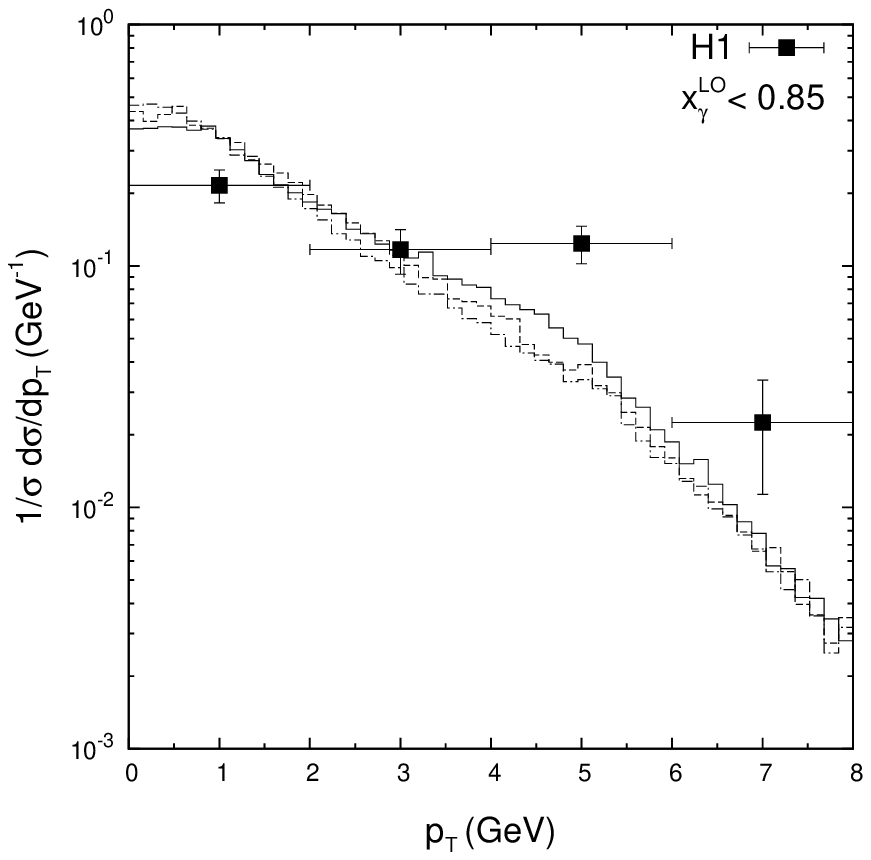, width = 8.1cm}
\epsfig{figure=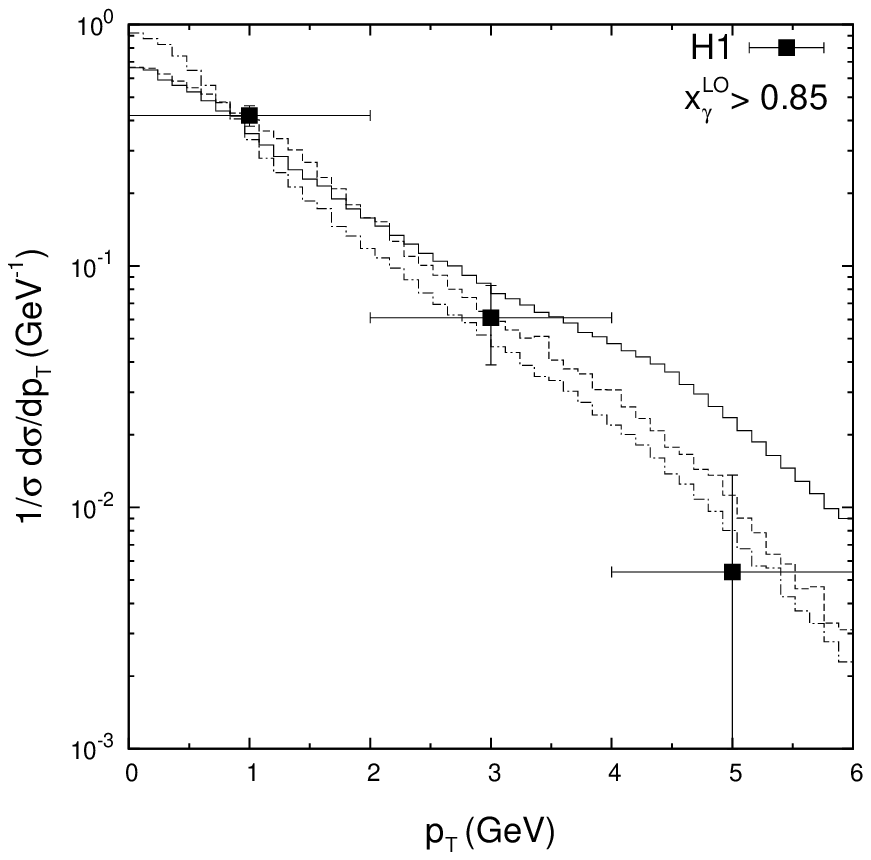, width = 8.1cm}
\caption{The normalized differential cross sections $1/\sigma\,d\sigma/dp_\perp$ 
for the prompt photon + jet production at HERA. Notation of all histograms 
is the same as in Figure~2. The experimental data are from H1~[5].}
\end{center}
\label{fig10}
\end{figure}

\begin{figure}
\begin{center}
\epsfig{figure=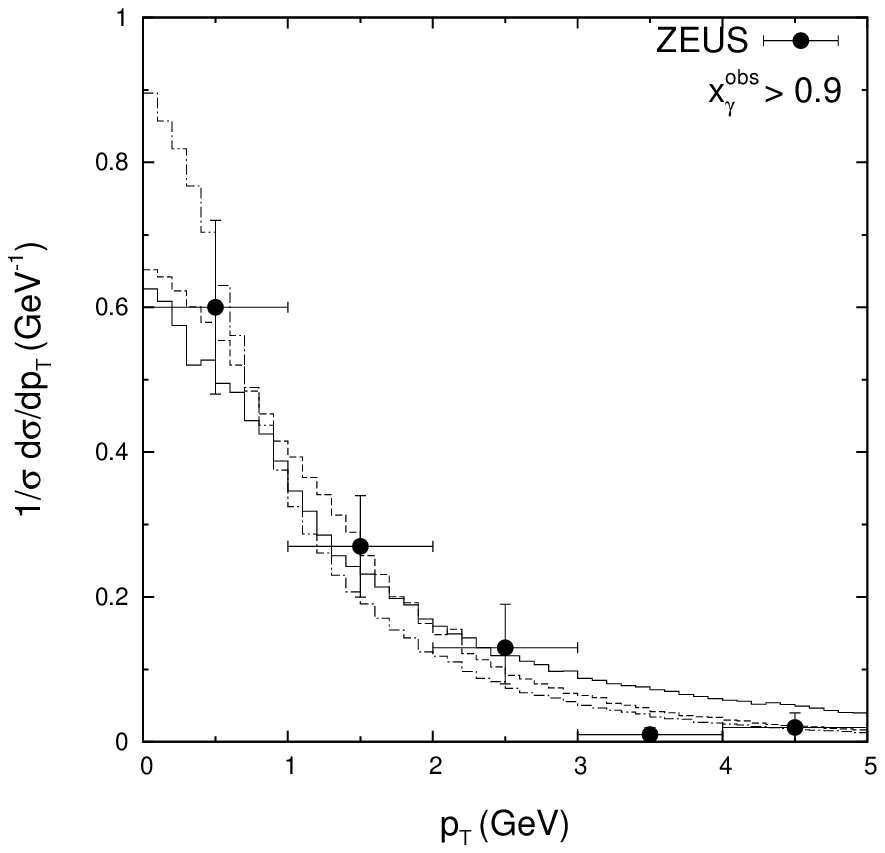, width = 8.1cm}
\epsfig{figure=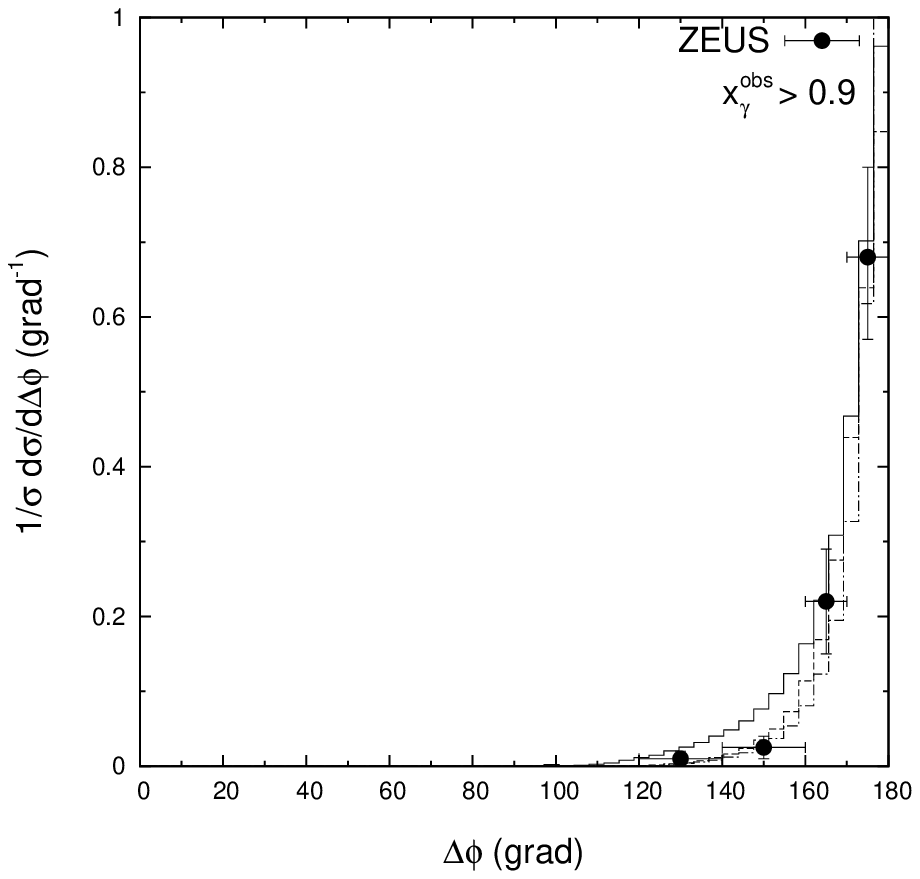, width = 8.1cm}
\caption{The normalized differential cross sections $1/\sigma\,d\sigma/dp_\perp$ and
$1/\sigma\,d\sigma/d\Delta \phi$ for the prompt photon + jet production at HERA. 
Notation of all histograms is the same as in Figure~1. The experimental data are from ZEUS~[3].}
\end{center}
\label{fig11}
\end{figure}

\end{document}